\documentclass[12pt]{article}
\usepackage{comment}
\usepackage{amsmath}
\usepackage{amssymb}
\usepackage{graphicx}
\usepackage{here}
\usepackage{subcaption}
\usepackage{url}
\usepackage[sort&compress, numbers, merge]{natbib}
\usepackage{physics}% Added by SK
\usepackage[compat=1.1.0]{tikz-feynhand}
\usepackage{slashed}
\usepackage{mathtools}

\usepackage{todonotes}

\numberwithin{equation}{section}

\usepackage[colorlinks=true, linkcolor=blue, citecolor=blue,
urlcolor=black]{hyperref}

\begin{document}
\def\ps{\mathbf{p}}
\def\PS{\mathbf{P}}
\baselineskip 0.6cm
\def\simgt{\mathrel{\lower2.5pt\vbox{\lineskip=0pt\baselineskip=0pt
           \hbox{$>$}\hbox{$\sim$}}}}
\def\simlt{\mathrel{\lower2.5pt\vbox{\lineskip=0pt\baselineskip=0pt
           \hbox{$<$}\hbox{$\sim$}}}}
\def\simprop{\mathrel{\lower3.0pt\vbox{\lineskip=1.0pt\baselineskip=0pt
             \hbox{$\propto$}\hbox{$\sim$}}}}
\def\tr{\mathop{\rm tr}}
\def\SU{\mathop{\rm SU}}

\begin{titlepage}

\begin{flushright}
IPMU21-0008
\end{flushright}

\vskip 1.1cm

\begin{center}

{\Large \bf 
On Stability of Fermionic Superconducting Current in Cosmic String
}

\vskip 1.2cm
Masahiro Ibe$^{a,b}$, 
Shin Kobayashi$^{a}$, 
Yuhei Nakayama$^{a}$ and
Satoshi Shirai$^{b}$
\vskip 0.5cm

{\it

$^a$ {ICRR, The University of Tokyo, Kashiwa, Chiba 277-8582, Japan}

$^b$ {Kavli Institute for the Physics and Mathematics of the Universe
 (WPI), \\The University of Tokyo Institutes for Advanced Study, \\ The
 University of Tokyo, Kashiwa 277-8583, Japan}
}

\vskip 1.0cm

\abstract{
Recently, the chiral superconductivity of the cosmic string in the axion model has gathered attention. The superconductive nature can alter the standard understanding of the cosmology of the axion model. For example, a string loop with a sizable superconducting current can become a stable configuration, which is called a Vorton. The superconductive nature 
can also affect the cosmological evolution of the string network.  In this paper, we study the stability of the superconducting current in the string. 
We find the superconductivity is indeed stable for a straight string or infinitely small string core size, even if the carrier particles are unstable in the vacuum. 
However we also find that the carrier particle decays in a curved string in typical axion models, if the carrier particles are unstable in the vacuum.
Accordingly, the lifetime of the Vorton is not far from that of the carrier particle
in the vacuum.
}

\end{center}
\end{titlepage}

\section{Introduction}
Topological defects associated with phase transitions in the early Universe have been the subject of intense study (see e.g., Ref.~\cite{Vilenkin:2000jqa}). Cosmic string is such a topological defect associated with U$(1)$ symmetry breaking.
One of the most important examples is the cosmic string in the axion model~\cite{Peccei:1977hh,Peccei:1977hh,Weinberg:1977ma,Wilczek:1977pj}, which solves the strong CP problem by utilizing the Nambu-Goldstone boson accompanying the breaking of the global U$(1)$ Peccei-Quinn (PQ) symmetry.
As an interesting phenomenological feature of the cosmic string, it can exhibit the superconductivity~\cite{Witten:1984eb}. For example, when the symmetry breaking field couples to chiral fermions with non-vanishing U$(1)_\mathrm{QED}$ charges, the cosmic string shows superconductivity. The bosonic superconducting string is also possible if the cosmic string induces the spontaneous breaking of U$(1)_\mathrm{QED}$ inside the string while U$(1)_\mathrm{QED}$ is unbroken outside of the string.

Recently, the superconductivity of the cosmic strings has gathered renewed attention in the axion models~\cite{Fukuda:2020kym,Abe:2020ure,Agrawal:2020euj}. 
As discussed in Ref.~\cite{Fukuda:2020kym},
the axion string becomes the chiral superconductor in which the KSVZ fermions carry the current traveling in only one direction along the string in the KSVZ axion model~\cite{Kim:1979if,Shifman:1979if}.
Once such a chiral superconducting string forms a closed loop, it stops shrinking at some point when the kinetic energy of the carrier fermions becomes non-negligible compared to the weight of the string loop.
Such a stable configuration is called a  Vorton~\cite{Carter:1993wu,Brandenberger:1996zp,Martins:1998gb,Martins:1998th,Carter:1999an}.
If the Vorton has a lifetime
of cosmological timescale, it contributes to the dark matter density and may significantly affect the cosmology of the axion models.

In this paper, we study the stability of the fermion which carries the superconducting current on the string (see also Refs.~\cite{Barr:1987ij,Davis:1999ec,Jeannerot:2004bs} for earlier works). 
We find that the carrier particle immediately decays in a curved string in typical axion models.
Accordingly, the lifetime of the string loop is not far from that of the carrier particles in the vacuum.
Thus the Vorton cannot be long-lived.

The organization of this paper is as follows.
In Sec.~\ref{sec:Axion}, we briefly review the axion model and the associated superconducting string.
In Sec.~\ref{sec:decayonmodulation}, we discuss the decay rate of the charge carrier when the strings are curved. 
In Sec.~\ref{sec:vorton}, we discuss the fate of the Vorton.
The final section is devoted to our conclusions.

\section{Axion and Superconducting String}
\label{sec:Axion}
In this section, we briefly review the superconductivity of the global string associated with the PQ-symmetry breaking.
\subsection{KSVZ Axion and Global String}
Throughout this paper, we consider
the KSVZ axion model~\cite{Kim:1979if,Shifman:1979if}, which consists of a complex scalar field $\phi$ and
$N_f$ KSVZ fermions, $\psi$, of the fundamental representation of the SU$(3)_c$ gauge group of QCD.%
\footnote{
In this paper, we use the convention for the four-component fermion in Ref.~\cite{Dreiner:2008tw}.
}
We will take $N_f = 1$ for successful cosmology (see later discussion).
The U(1) PQ-symmetry is defined by the phase rotations,
\begin{align}
\label{eq:PQcharge}
    \phi \to \phi' = e^{i \alpha} \phi\ , \quad 
   \psi_L \to \psi_L' =  e^{-i\alpha}\psi_L\ , \quad 
   \psi_R \to \psi_R' =  \psi_R\ ,
\end{align}
where $\psi_{L,R} = P_{L,R} \psi$ and $\alpha$ is a rotation parameter.
 The PQ-symmetry allows couplings between $\phi$ and $\psi$, 
\begin{align}
\label{eq:Yukawa}
   &\mathcal{L}= \partial_\mu \phi^* \partial^\mu \phi 
   + \bar{\psi}i\gamma^\mu\partial_\mu \psi - V(\phi) 
   + y_\psi\phi \bar{\psi} P_L \psi    + h.c.\ , \\
   &
   \label{eq:potential}
   V(\phi)= \frac{\lambda}{4} (|\phi|^2 - v_{\mathrm{PQ}}^2)^2 \ ,
\end{align}
where $y_\psi (> 0)$ is a Yukawa coupling constant, $\lambda (> 0)$ 
the quartic coupling constant, and $v_{\mathrm{PQ}}$ is a parameter with a mass dimension one.

At the vacuum, the PQ-symmetry is spontaneously broken by the VEV of $\phi$,
\begin{align}
\langle \phi \rangle = v_{\mathrm{PQ}}\ ,
\end{align}
where we take $v_{\mathrm{PQ}} > 0$.
From the Yukawa coupling, the KSVZ fermion obtains a mass,
\begin{align}
m_\psi = y_\psi v_{\mathrm{PQ}}\ .
\end{align}
In the broken phase, the PQ-symmetry is realized as a shift symmetry of the axion, $a$,
\begin{align}
    a/f_{\mathrm{PQ}} \to a'/f_{\mathrm{PQ}} = a/f_{\mathrm{PQ}} + \alpha , \quad 
    \psi_L \to \psi_L' =  e^{-i\alpha}\psi_L\ , \quad 
   \psi_R \to \psi_R' =  \psi_R\ , 
\end{align}
where the axion resides in $\phi$ as,
\begin{align}
\phi = v_{\mathrm{PQ}}\,e^{ia/f_{\mathrm{PQ}}}\ , \quad f_{\mathrm{PQ}} = \sqrt{2} v_{\mathrm{PQ}}\ .
\end{align}
Due to the QCD anomaly, the axion obtains a non-trivial scalar potential, and the effective $\theta$-angle of the QCD is erased at the minimum of the axion potential.

Associated with the PQ-breaking,
there are the global cosmic string configurations which have non-trivial topological numbers in $\pi_1(\mathrm{U}(1))$  (see e.g., Ref.~\cite{Vilenkin:2000jqa}).
For a straight cosmic string, the string configuration is given by,
\begin{align}
\label{eq:string}
    \phi = v_{\mathrm{PQ}}h(\rho) e^{in_w \varphi}\ ,
\end{align}
where ($\rho, \varphi,z$) is the cylindrical coordinate along the straight string in the $z$-direction.
The integer $n_w \in \pi_1(\mathrm{U}(1))$ is the winding number.
For a scalar potential in Eq.\,\eqref{eq:potential},
the field equation of $h$ is given by,
\begin{align}
\label{eq:h}
 h''(\rho) + \frac{h'(\rho)}{\rho} - \frac{n_w^2}{\rho^2} h(\rho) - \frac{1}{2}m_\phi^2(h(\rho)^2-1)h(\rho)  = 0\ ,
\end{align}
with boundary conditions,
\begin{align}
&h(\rho)\propto \rho^{|n_{w}|} \ , \quad (\rho\to 0)\ ,\\
&h(\rho)\to 1\ , 
\quad \,\,\,\,\,\,\,(\rho \to \infty)\ .
\end{align}
Here, $m_\phi$ denotes the mass of the modulus component of $\phi$, $m_{\phi} = \sqrt{\lambda}v_{\mathrm{PQ}}$.%
\footnote{Asymptotically, $h(\rho) = 
%1 - \order{(m_\phi\rho)^{-2}}
\sqrt{1 - 2 n_w^2/(m_\phi\rho )^2}
$ for $m_\phi\rho \gg 1$.}

Unlike the Abrikosov-Nielsen-Olesen
(ANO) local string~\cite{Nielsen:1973cs}, the tension of the global string is logarithmically divergent
since the angular gradient of $\phi$ is not compensated by the gauge field.
Nonetheless, those strings are expected to be formed at the phase transition in the early Universe, 
where the divergence is cut off by a typical distance between the strings of the order of the Hubble length, $H^{-1}$.
That is, the string tension $\mu$ at the formation is roughly given by,
\begin{align}
    \mu \sim 2\pi v_{\mathrm{PQ}}^2 \log \frac{m_{\phi}}{H}\ .
\end{align}

We assume that the PQ-breaking takes place after the end of inflation.
In the following analysis, we take $N_f = 1$, 
otherwise, the KSVZ axion potential has a discrete $\mathrm{Z}_{N_f} (N_f > 1)$ symmetry which is broken by the VEV of the axion. 
Thus, the model with $N_f > 1$ suffers from the domain-wall problem, while the model with $N_f = 1$ is free from this problem.%
\footnote{If we allow flavor dependent PQ charges of $\psi$'s, it is possible to avoid the domain wall number for $N_f > 1$.}

In the model with $N_f= 1$, the axion dark matter density is dominated by the axions produced by the decay of the string-wall network at the QCD phase transition
in the case that the PQ breaking takes place after inflation.
The simulations suggest the axion abundance exceeds the observed dark matter density for $f_{\mathrm{PQ}} >\order{10^{11}}$\,GeV~\cite{Hiramatsu:2010yn,Hiramatsu:2012gg,Gorghetto:2018myk,Gorghetto:2020qws}.
To avoid cosmological and astrophysical constraints, we consider the range of the axion decay constant in,
\begin{align}
     f_{\mathrm{PQ}} = 10^8- 10^{11}\,\mathrm{GeV}\ ,
\end{align}
where the lower bound comes from the astrophysical constraints~\cite{Raffelt:2006cw,Chang:2018rso,Irastorza:2018dyq,Hamaguchi:2018oqw}.

In the case that the PQ-breaking takes place after inflation, the KSVZ fermions are in the thermal-equilibrium.
Since they become heavy due to the PQ-breaking and get the mass $y_\psi v_{\mathrm{PQ}}$,  they need to decay into the SM particles to avoid the cosmological problems.
The KSVZ fermions can decay well before the Big-Bang Nucleosynthesis if, for example, $\psi_R$ has the same quantum number as the down-type quark, ${d}_R$, in the SM.%
\footnote{As defined in Eq.\,\eqref{eq:PQcharge}, $\psi_R$ is neutral under the PQ symmetry.}
In this case, the decay operator is given by,
\begin{align}
\label{eq:decay}
    \mathcal{O}_D = y_{D}H_\mathrm{SM}\bar{\psi}_Rq_{L} + h.c.\ ,
\end{align}
where $H_\mathrm{SM}$ and $q_L$ are the Higgs and the quark doublets in the SM. 
The corresponding decay rate is given by,
\begin{align}
    \Gamma_D = \frac{|y_D|^2}{16\pi} m_\psi \ . \label{eq:free_decay}
\end{align}
Here, we consider the decay rate into only one generation of the quark doublets for simplicity.

Note that $\psi$ has the same gauge charges with ${d}_R$.
In this case, the KSVZ fermion has the QED charge $q_\psi = -1/3$.
We may instead assign $\psi_R$ the same quantum number with the up-type quarks, ${u}_R$. In this case, the QED charge of $\psi$ is given by $q_\psi = 2/3$, although the change of the charge does not affect the following arguments.
\subsection{Superconductivity of Axion String}
Around an infinitely long string, the Dirac equation of the fermion coupling to the string
has the zero mode solutions which are normalizable in the transverse plane of the string~\cite{Weinberg:1981eu,Jackiw:1981ee}. 
The normalizable fermion zero mode can propagate along the string as a massless mode.
When the fermion zero mode has non-trivial gauge charges, it carries the superconducting current~\cite{Witten:1984eb}.
Recently, Ref.~\cite{Fukuda:2020kym} revisited the string superconductivity in the KSVZ axion model.
It showed that the axion strings generically exhibit the chiral superconductivity. 
We here briefly review the fermion zero modes and the superconductivity of the string.

We consider the Dirac equation around the straight cosmic string along the $z$-axis,
\begin{align}
    \left[i\gamma^\mu\partial_\mu - m_\psi h(\rho)
    \left(
    e^{i\varphi}P_L + e^{-i\varphi}P_R
    \right)
    \right]\psi= 0 \ .
\end{align}
where $\rho= \sqrt{x^2 + y^2}$ and $\varphi$ is the azimuthal angle on the $(x,y)$ plane. 
Here, we assume the winding number of the string is $n_w = 1$.
For now, let us assume that $\psi$ depends only on the transverse coordinate $(x,y)$. Then, the Dirac equation is reduced to,
\begin{align}
    &i \gamma^1\left(\partial_1 + i(i\gamma^1\gamma^2)\partial_2\right)\psi_{L} = m_\psi h(\rho)e^{-i\varphi} \psi_{R}\ , \\
    &i \gamma^1
    \left(\partial_1 +i(i\gamma^1\gamma^2)\partial_2\right)\psi_{R} = m_\psi h(\rho)e^{i\varphi} \psi_{L}\ .
\end{align}
The explicit solution which is normalizable in the transverse plane is given by \cite{Jackiw:1981ee},
\begin{align}
\label{eq:zeromode}
    \psi^0(x,y) = \mathcal{N} \eta
    \exp(-\int_0^\rho m_\psi h(\rho')d\rho')\ , \quad 
    \eta =   \left[\begin{array}{c}
         0  \\
         1 \\
         i \\
         0
    \end{array}
    \right]\ ,
\end{align}
which satisfies%
\footnote{It is convenient to note $\partial_1 \pm i \partial_2 = e^{\pm i\varphi } (\partial_\rho \pm i \rho^{-1} \partial_\varphi)$.}
\begin{align}
    i\gamma^1\gamma^2\psi_{L}^0 = - \psi_{L}^0\ , \quad 
    i\gamma^1\gamma^2\psi_{R}^0 = \psi_{R}^0\ , \quad
    \psi_{R}^0 = -i \gamma^1 \psi_{L}^0\ ,
    \quad \partial_\varphi \psi_{L,R}^0= 0 \ .
\end{align}
The normalization constant $\mathcal{N}$ will be determined later.
The zero-mode configuration in Eq.\,\eqref{eq:zeromode} is localized at around the core of the cosmic string with a finite size of $\mathcal{O}(m_\psi^{-1})$.

Interestingly, the normalizable transverse zero modes lead to the ``massless" propagation modes along the cosmic string.
Let us the assume, 
\begin{align}
\label{eq:massless}
    \psi^0(t,x,y,z) = \alpha(t,z)\psi^0(x,y)\ .
\end{align}
then, the Dirac equation in the four-dimensional spacetime is reduced to,
\begin{align}
    (\gamma^0\partial_0 + \gamma^3 \partial_3)\alpha(t,z)\eta = 0 \ , 
\end{align}
or 
\begin{align}
    (\partial_0 +  \partial_3)\alpha(t,z) = 0 \ , 
\end{align}
since $\gamma^0\gamma^3 \eta = \eta$.
The general solution of this equation,
\begin{align}
    \alpha(t,z) = \alpha(t-z) \ ,
\end{align}
corresponds to the ``massless" chiral fermion which moves at the speed of light in the positive $z$-direction.
Note that both the particle $\psi^0$ and the anti-particle, $\psi^{0\,c}=i\gamma_2\psi^{0*}$, move in the same direction along the string.
They also have the same chirality in the longitudinal two-dimensional spacetime, i.e., $\gamma^0\gamma^3 \psi^{0(c)} =  \psi^{0(c)}$. 

When the straight string is placed in a constant electric field in the $z$-direction, $E_z$, the fermion zero modes are induced on the string. 
The resulting current on the string grows in the applied time $\mathit{\Delta}t$ as~\cite{Witten:1984eb},
\begin{align}
J = \frac{q_\psi^2}{2\pi} E_z \mathit{\Delta}t \ ,
\end{align}
which is a characteristic of a superconducting wire.
This current is persistent and remains even after the electric field is turned off.%
\footnote{In Ref.\,\cite{Witten:1984eb}, it is considered the vector-like charge carrier on the string.
For detailed discussion on the chiral nature of the  superconducting string, see e.g., Refs.\,\cite{Callan:1984sa,Kaplan:1987kh,Naculich:1987ci}.}

\section{Stability of Fermion Zero Mode in Curved String}
\label{sec:decayonmodulation}
In the above discussion, we considered the straight cosmic string.
In general, however, the cosmic strings are curved.
Macroscopically, cosmic strings have a curvature radius of $\order{H^{-1}}$ (see e.g., Refs.~\cite{Hiramatsu:2010yn,Hiramatsu:2012gg}).
They also have microscopic curvatures induced by the thermal fluctuations.
In this section, we consider the decay of the fermion zero mode in a curved string.
\subsection{Schematic Picture of Zero Mode Decay}
\label{sec:classical_escape}
\begin{figure}
    \centering
    \includegraphics[width=0.8\linewidth]{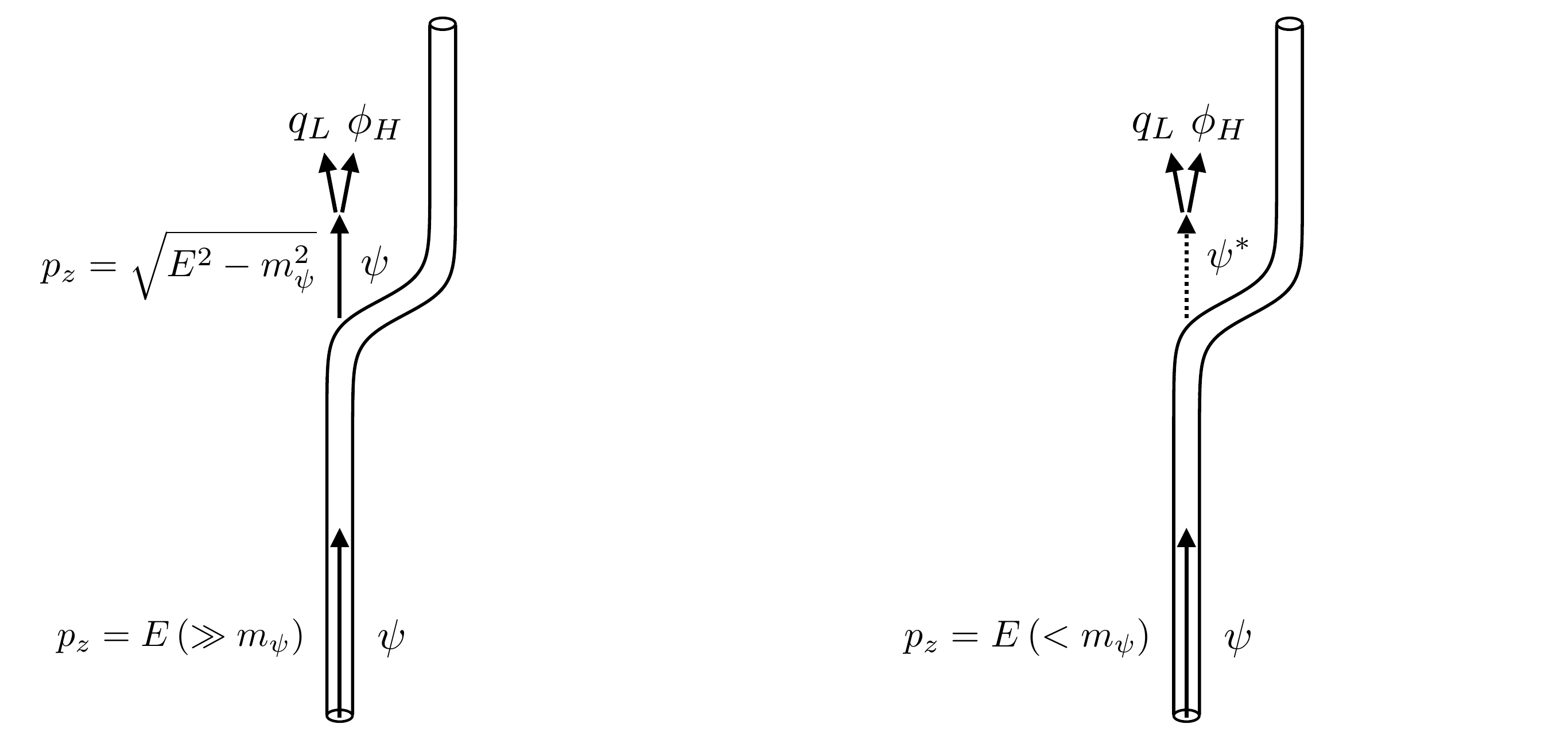}
    \caption{Schematic pictures of the decay of the fermion zero mode in the curved string. The curvature radius of the curve is $\order{m_\phi^{-1}}$. For the zero mode with $\abs{p_z} \gg m_\psi$, it escapes from the string at the curve and decays via the decay operator.
    For the zero mode with $\abs{p_z} < m_\psi$, it decays through the quantum tunneling.}
    \label{fig:curved}
\end{figure}

For a zero mode in a straight cosmic string, its decay is prevented by the energy conservation and the longitudinal momentum conservation.
The initial energy and the momentum in the $z$-direction are $E_i=p_i^z = E$, respectively. Thus, the energy and the longitudinal momentum conservation leads to, $E_f = p_f^z = E$, and hence, the phase space of the transverse momentum of the final state is vanishing.
For a zero mode in a curved string with finite string core size, on the other hand, it is expected that the fermion zero mode decays into the SM particles through the decay operator since the longitudinal momentum conservation is broken by the curvature of the string~(see Refs.~\cite{Davis:1988ip,Jeannerot:2004bs} for related works).

For example, let us consider a curved string with its curvature radius is comparable to the string core size, i.e., $\order{m_\phi^{-1}}$ (see Fig.\,\ref{fig:curved}).
In this case, the zero mode with $\abs{p_z} \gg m_\psi$ simply goes out of the string and behaves as a massive free fermion~\cite{Witten:1984eb} (see the Appendix~\ref{sec:classical}).
Thus, it immediately decays into the SM particles through the decay operator.
For the zero mode with $\abs{p_z} < m_\psi$, on the other hand, it does not go out of the string due to the energy conservation.
Even in this case, it is expected that the fermionic zero mode decays into the SM particles through the off-shell fermion, i.e., the quantum tunneling.

Let us emphasize that the fermion zero mode is stable even on a curved string when the core radius of the string is vanishing, i.e., $m_\phi \to \infty$.
In this limit, the straight string picture is valid at any point on the string even if it is curved with a finite curvature radius.%
\footnote{In other words, the fermion zero mode does not have injection momentum onto the potential wall made by the curve of the string for $m_\phi\to\infty$.}
Thus, the fermion zero mode is trapped inside the two-dimensional spacetime along the string.
Hence, even the fermion zero mode with $E\gg m_\psi$ cannot escape from the string and moves along the string.

\subsection{Decay of Fermion Zero Mode in Curved String} 
\label{sec:decay_master}
To discuss the decay rate of the zero mode in a curved string, let us consider a slight modulation on a straight string along the $z$-axis. 
We consider a modulation in the $y$-direction in which the position of the string center is given by
\begin{align}
\label{eq:curve}
    (x,y,z) = \left(0, f(z), z\right)\ .
\end{align}
We consider that the maximal value of $|f|$ is much smaller than the string core size $m_\phi^{-1}$, so that the modulation can be treated perturbatively.
In the following, we consider a time-independent modulation.
The curve in Eq.\,\eqref{eq:curve} modulates the string profile function $h(\rho)$,
\begin{align}
\label{eq:modulation}
    h(\rho) \to h(\rho) + \delta h(\rho,\varphi,z)\ , \quad
    \delta h(\rho,\varphi,z) = \frac{y}{\rho}\frac{d h(\rho)}{d\rho}\times f(z)\ ,
\end{align}
for $|f|\ll m_\phi^{-1}$.
The modulation couples to the fermion via the Yukawa coupling,
\begin{align}
\label{eq:yukawa_modulation}
    \mathcal{O}_M = m_\psi \delta h(\rho,\varphi,z) \bar{\psi}(e^{i\varphi}P_L+e^{-i\varphi}P_R)\psi \ ,
\end{align}
which originates from Eq.\,\eqref{eq:Yukawa}.

The quantized KSVZ fermion field in the presence of a straight string along the $z$-axis with a length $L_{\mathrm{str}}$ and a winding number $n_w = 1$ is given by
\begin{align}
    \hat\psi =& \frac{1}{\sqrt{L_{\mathrm{str}}}}\sum_{n>0}
    \left(e^{-iE_n(t-z)} u(\rho)\, \hat{b}^{0}_n + e^{iE_n(t-z)} v(\rho)\, \hat{d}^{0}_n{}^\dagger
    \right) \cr
    &+\sum 
(\mbox{bounded massive modes})
+
\sum (\mbox{unbounded modes}) \ .
\end{align}
The first line denotes the zero-mode contribution of energy $E_n = 2\pi n/L_{\mathrm{str}}$ $(n>0)$.
The wave functions $u(\rho)$ and $v(\rho)$are given by,
\begin{align}
\label{eq:wavefunction}
    u(\rho) = \mathcal{N} \eta \exp\left(-\int_0^\rho m_\psi h(\rho')d\rho'\right)\ , \quad
    v(\rho) = i\gamma_2 u(\rho)^*
    \ ,\quad \int dxdy |u(\rho)|^2 = 1 \ ,
\end{align}
where $h(\rho)$ is the profile function of a straight string in Eq.\,\eqref{eq:h}.
The normalization constant  $\mathcal{N}$ is fixed by the third condition in Eq.\,\eqref{eq:wavefunction}.%
\footnote{The normalization constant is approximately 
given by 
$\mathcal{N} \simeq m_\psi/\sqrt{{2\pi}}$ for $m_\phi\gg m_\psi$ .
}
The second and the third terms represent the massive bounded modes in the string and the unbounded continuous modes, respectively~\cite{Davis:1999ec,Ringeval:2000kz,Ringeval:2001xd}.
The creation operators of the zero modes satisfy,
\begin{align}
\label{eq:quantized1}
    \{\hat{b}^0_n,\hat{b}^{0\dagger}_{n'}\} = \delta_{nn'}\ , \quad \{\hat{d}^0_n,\hat{d}^{0\dagger}_{n'}\} = \delta_{nn'}\ .
\end{align}
The unbounded continuous mode asymptotically behaves as a massive free particle with a mass $m_\psi$ away from the string.

In the limit of $L_\mathrm{str}\to \infty$, the quantized field becomes
\begin{align}
\label{eq:quantized2}
    \hat\psi =& \int_0^\infty \frac{dE}{2\pi}
    \left(e^{-iE(t-z)} u(\rho)\, \hat{b}^{0}(E) + e^{iE(t-z)} v(\rho)\, \hat{d}^{0}(E){}^\dagger
    \right) \cr
    &+\int 
(\mbox{bounded massive modes})
+ \int (\mbox{unbounded modes}) \ ,
\end{align}
with the creation/annihilation operators,
\begin{align}
\label{eq:normalization}
    \{\hat{b}^0(E),\hat{b}^{0}(E')^\dagger\} = (2\pi)\delta(E-E')\ , \quad  \{\hat{d}^0(E),\hat{d}^{0}(E')^\dagger\} = (2\pi)\delta(E-E')\ .
\end{align}
Here, $\hat{b}^0(E)$ and $\hat{d}^0(E)$ correspond to $\sqrt{L_{\mathrm{str}}}\,\hat{b}^0_n$ and $\sqrt{L_{\mathrm{str}}}\,\hat{d}^0_n$, respectively.

%----------------------------
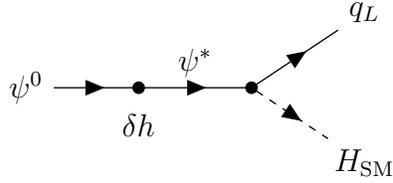
\begin{figure}[tpb]    
\begin{center}      
\begin{tikzpicture} %tikzpicture
\begin{feynhand}    %feynhand
    \vertex [particle] (i1) at (-1.5,3) {$\psi^0$};
 \vertex [particle] (f1) at (3,4) {$q_L$};
\vertex [particle] (f2) at (3,2) {$H_\mathrm{SM}$};
\vertex [particle] (i2) at (0,2.5) {$\delta h$};
    \vertex [dot] (w1) at (0,3) {};
     \vertex [dot](w2) at (1.5,3){};
    \propag [fermion] (i1) to (w1);
   \propag [fermion] (w1) to  [edge label = $\psi^{*}$] (w2) ;
   \propag [fermion] (w2) to (f1);
    \propag [chasca] (w2) to (f2);
\end{feynhand}
\end{tikzpicture}
\caption{The Feynman diagram corresponding to the leading order contribution to the zero mode decay into the quark and the Higgs in Eq.\,\eqref{eq:Born}. }
\label{fig:Diagram}
\end{center}
\end{figure}
%----------------------------

Now, let us calculate the matrix element of the decay process of the fermion zero mode by using the quantized field in Eq.\,\eqref{eq:quantized2}.
In the presence of the modulation, the matrix element for the decay process is given by
\begin{align}
\label{eq:amplitude0}
\hat{T} = \langle 0| \hat{a}_q \hat{a}_H Te^{i\int d^4x\left[\mathcal{O}_M+\mathcal{O}_D\right] } \hat{b}^0(E)^{\dagger}|0\rangle \ .
\end{align}
Here, $\hat{a}_{q}$ and $\hat{a}_H$ are the annihilation operators of the quark and the Higgs doublets, respectively.
The state $|0\rangle$ denotes the ground state with a global string with $n_w = 1$.
In the Born approximation (Fig.\,\ref{fig:Diagram}), the amplitude becomes,
\begin{align} 
\label{eq:Born}
\hat{T} \simeq -y_D^*m_\psi \bar{u}_{\mathrm{q}}P_R
\int d^4x\int d^4x' \delta h(x) 
\langle 0|
T\psi(x')\bar{\psi}(x)|0\rangle P(\varphi)u(\rho)e^{-iE(t-z)}e^{ip_fx'}\ .
\end{align}
Here $\bar{u}_q$ is the wave function of $q_L$ in the final state, $P(\varphi) =  e^{i\varphi}P_L + e^{-i\varphi}P_R$, and $p_f$ denotes the sum of the final state four-dimensional momenta.

In the presence of the string, the exact form of the propagator of $\psi$ is not known even for $\delta h = 0$.
In the present analysis, we only take account of the unbounded fermion contributions to the propagator and approximate it by that of the free fermion,
\begin{align}
\label{eq:propagator}
    \langle 0|
T\psi(x)\bar{\psi}(x')|0\rangle
\simeq 
\int\frac{d^4p}{i(2\pi)^4}
\frac{m_\psi+\slashed{p}}{m_\psi^2 - p^2 + i \epsilon}e^{-ip(x-x')}\ .
\end{align}
Under this approximation, the matrix element is reduced to,
\begin{align}
    \hat{T} \simeq -y^*_D m_\psi \bar{u}_{\mathrm{q}}P_R
\int d^4x\int d^4x' \delta h(x) 
\int\frac{d^4p}{i(2\pi)^4}
\frac{m_\psi+\slashed{p}}{m_\psi^2 - p^2 + i \epsilon}e^{ip(x-x')}
 P(\varphi)u(\rho)e^{-iE(t-z)}e^{ip_fx'}\ .
 \end{align}
By integrating over $x'$, $p$ and $t$, we obtain,
\begin{align}
\label{eq:hatT}
    \hat{T} \simeq i(2\pi)\delta(E-E_f)y_D^*m_\psi \bar{u}_{\mathrm{q}}P_R
\int d^3x \delta h(x) 
\frac{e^{-i\varphi}m_\psi+e^{+i\varphi}\slashed{p}_f}{m_\psi^2 - p_f^2 + i \epsilon} u(\rho)e^{iEz}e^{-i\mathbf{p}_f\cdot\mathbf{x}}\ .
\end{align}
Incidentally, the off-shell fermion zero modes do not contribute to the decay rate in the Born approximation since
\begin{align}
\left.    \langle 0|
T\psi(x)\bar{\psi}(x')|0\rangle
\right|_{\mbox{\footnotesize{zero mode}}}\propto 
\left[
\begin{array}{cccc}
     0&0&0&0  \\
     -i&0&0&i  \\
     1&0&0&-1  \\
     0&0&0&0 
\end{array}
\right]\ ,
\end{align}
and hence, $\left.    \langle 0|
T\psi(x)\bar{\psi}(x')|0\rangle
\right|_{\mbox{\footnotesize{zero mode}}}\times P(\varphi)u(\rho)=0$.

To specify the modulation, we consider a periodic curve with a period $L$, that is,
\begin{align}
\label{eq:sinuous}
    &f(z) = \sum_{n=-\infty}^\infty c_n e^{-i\frac{2\pi n}{L}z}\ , \\
    &c_n = \frac{1}{L}\int_{-L/2}^{L/2} dz e^{i\frac{2\pi n}{L}z}f(z) \ ,
\end{align}
with $c_{-n} = c_n^*$.%
\footnote{Strictly speaking, we need to assume $\delta h(z) \to 0$ for $|z| \to \infty$ to justify the amplitude in Eq.\,\eqref{eq:amplitude0}.
This assumption requires a dumping factor at $|z|\to \infty$.
However, since the decay rate should not depend on the shape at $|z|\to\infty$, we estimate the decay rate by using the periodic modulation as it is.
In the appendix~\ref{sec:SingleGaussian}, we confirm that the consistent decay rate is obtained for a non-periodic modulation with $\delta h(z)\to 0$ at $|z|\to \infty$.
}
In this case, the integration over $z$ results in,
\begin{align}
\label{eq:Tdxdy}
    \hat{T} 
\simeq& i(2\pi)^2\delta(E-E_f)
y_D^* m_\psi\bar{u}_{\mathrm{q}}P_R
\sum_{n}c_n\delta(E-p_{f}^z-k_n)
\cr
&\qquad\qquad\times \int d^2x \frac{e^{-i\varphi}m_\psi+e^{+i\varphi}\slashed{p}_f}{m_\psi^2 - p_f^2 + i \epsilon} \sin\varphi 
\frac{dh}{d\rho}u(\rho)e^{-i
\left.\mathbf{p}_f\cdot\mathbf{x}\right|_{\perp}}\ ,
\end{align}
where $k_n = 2\pi n/L$ $(n\in \mathbb{Z})$ and the subscript $\perp$ denotes the $(x,y)$ space components.
For each a modulation mode, the total invariant mass of the final state satisfies,
\begin{align}
    0 \le p_f^2 \le E^2-(E-k_n)^2 = 2 Ek_n -k_n^2 \ .
\end{align}
Thus, the decay rate of the high energy mode in a slowly curved string, $E\gg |k_n|$, gets contributions only from the modes with $n>0$.
The result also confirms that the decay rate is vanishing in a straight string, $n=0$, where $p_f^2 = 0$.

\begin{figure}[tbp]
\centering{\includegraphics[width=0.5\textwidth]{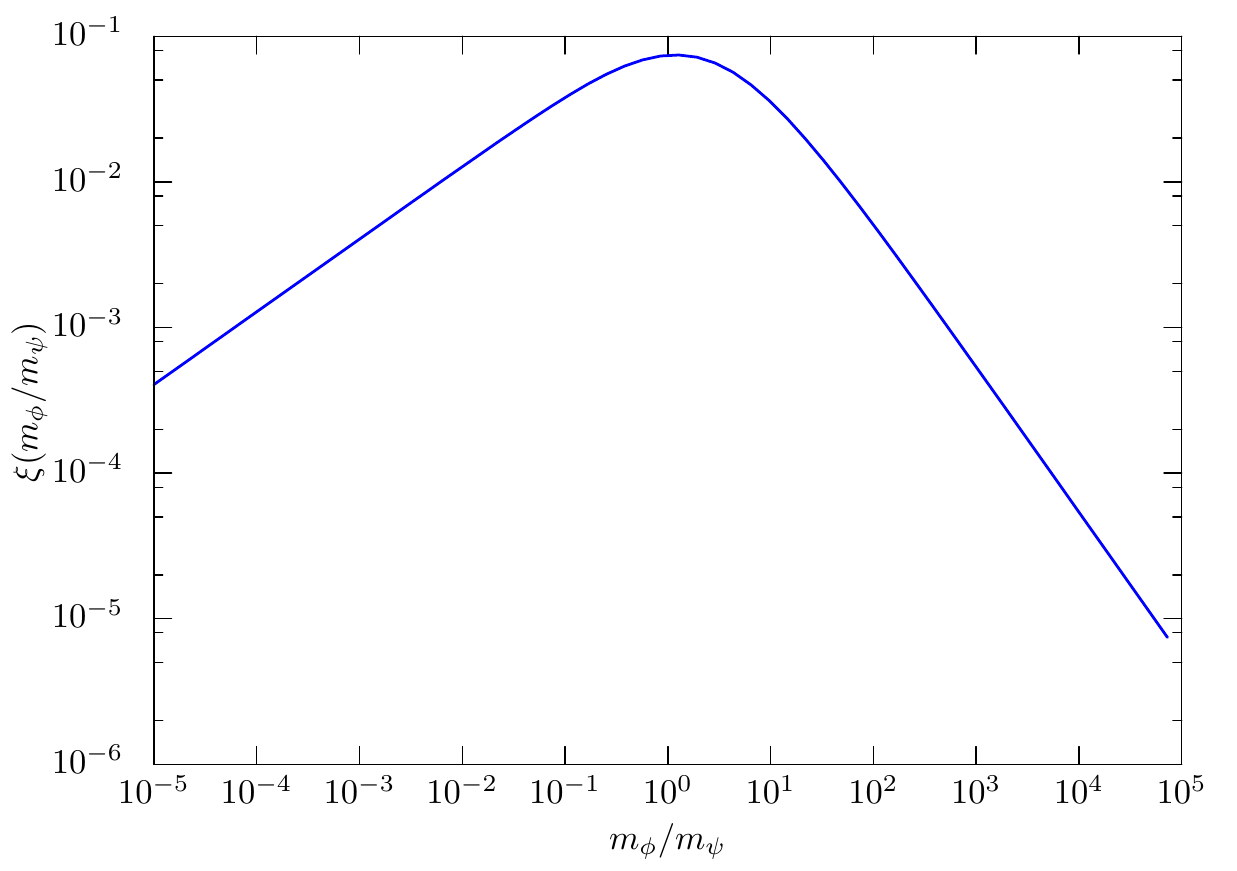}} 
\caption{
The numerical estimation of the function $\xi(m_\phi/m_\psi)$.
}
\label{fig:xir}
\end{figure}

Our main interest is the fate of the fermion zero modes with $E\lesssim v_{\mathrm{PQ}}$ in a slowly curved string, $k_n \ll m_\phi$.
For such a fermion, we find  $p_f^2 < m_\psi^2$, $\slashed{p}_{f}\eta \ll m_\psi \eta$, and hence, the amplitude is reduced to
\begin{align}
\label{eq:amplitude}
\hat{T} \simeq i(2\pi)^2\delta(E-E_f)
y_D^* \bar{u}_{\mathrm{q}}P_R
\sum_{n>0}c_n\delta(E-p_{f}^z-k_n)
\int d^2x e^{-i\varphi} \sin\varphi 
\frac{dh}{d\rho}u(\rho)e^{-i
\left.\mathbf{p}_f\cdot\mathbf{x}\right|_{\perp}}\ .
\end{align}
The integrand is highly suppressed for $\rho \gg m_{\phi,\psi}^{-1}$.
In this region, 
 $\left.\mathbf{p}_f\cdot\mathbf{x}\right|_{\perp} \ll 1$,
 and hence, 
\begin{align}
    \hat{T}\simeq (2\pi)^3\delta(E-E_f)y_D^* 
\bar{u}_{\mathrm{q}}P_R
\sum_{n>0}c_n\delta(E-p_{f}^z-k_n)
\xi\left(\frac{m_\phi}{m_\psi}\right)
\eta\ ,
\end{align}
where $\xi(m_\phi/m_\psi)$ is defined by,
\begin{align}
    \xi\left(\frac{m_\phi}{m_\psi}\right)= \xi\left({m_\phi},{m_\psi}\right) = \frac{i\mathcal{N}}{2\pi} \int d^2x e^{-i\varphi} \sin\varphi 
\frac{dh}{d\rho}
 \exp\left(-\int_0^\rho m_\psi h(\rho')d\rho'\right)\ .
\end{align}
Asymptotically $\xi(m_\phi/m_\psi \to \infty) \to 0.54 \times m_\psi/m_\phi$ and $\xi(m_\phi/m_\psi \to 0) \to 0.13 \sqrt{m_\phi/m_\psi}$ (see Fig.~\ref{fig:xir}).%
\footnote{Here, 
the numerical coefficients originate from
$0.13 = \sqrt{h'(0)}/(2\sqrt{2\pi}) $ and $0.54 = \int_0^\infty y h' dy/2\sqrt{\pi}$ with $y=\rho m_{\phi}$.}
The suppression of the amplitude for $m_\phi/m_\psi \gg 1 $ is reasonable since the straight string picture becomes valid at any point on the string in this limit.

By dividing the squared matrix element by the total time and the total length of the string,%
\footnote{The total time and the length are given by
$T_{\mathrm{interval}} = (2\pi)
\delta(E)|_{E = 0}$ and 
$L_{\mathrm{str}}=(2\pi)\delta(p_z)|_{p_z = 0} $, respectively.}
we obtain the decay rate of one fermion zero mode with energy $E$,
\begin{align}
\label{eq:decayrate_integral}
\Gamma(E) \simeq 
|y_D|^2 \xi^2\left(\frac{m_\phi}{m_\psi}\right)
 \sum_{n>0}|c_n|^2\!\int\!\!\! \frac{d^3\mathbf{p}_q}{(2\pi)^3 2p_q^0}
\frac{d^3\mathbf{p}_H}{(2\pi)^3 2p_H^0}
(2\pi)^4\delta(E-E_f)\delta(E-p_{f}^z-k_n) (p_q^0 - p_q^z)\ .
\end{align}
Here, $\mathbf{p}_{q,H}$ are the three momenta of the quark and the Higgs doublets. We neglect the masses of the quarks and Higgs.
Throughout this paper, we consider the decay rate of the in-flight fermion.
This is appropriate since we are interested in the decay rate of the fermion measured from outside of the string. 
The phase space integration (see the appendix~\ref{sec:phasespace}) results in
\begin{align}
\label{eq:decayrate_master}
   \Gamma(E) \simeq 
\frac{1}{24}|y_D|^2 \xi^2\left(\frac{m_\phi}{m_\psi}\right)E^3 \sum_{n>0}^{k_n <2 E}|c_n|^2 \mathcal{F}(k_n/E)\ , 
\end{align}
where 
\begin{align}
\mathcal{F}(x) = x^2(1-x/2)\ .
\end{align}

Let us comment on the case of a sharply curved string, which 
can appear as kinks and cusps in the cosmic string (see e.g., Ref.~\cite{Gouttenoire:2019kij}).
In this case, $p_f^2$ can be of $\order{m_\psi^2}$, and hence, the contribution of the pole of the propagator in Eq.\,\eqref{eq:Tdxdy} is significant.
This contribution corresponds to the process in the left panel of Fig.\,\ref{fig:curved} in which the fermion zero mode escapes from the string and decays.
The decay rate of such a process is also suppressed for a large $m_\phi$ due to the factor of $dh/d\rho$ in the integrand of Eq.\,\eqref{eq:Tdxdy}.
This suppression is consistent with the observation that the fermion zero mode does not decay in a straight string, because the string looks like a straight one at any point in the limit of $m_\phi \to \infty$.

In this analysis, we do not take into account the contributions of the bounded massive modes to the fermion propagator~\cite{Davis:1999ec,Ringeval:2000kz,Ringeval:2001xd}.
The fermion zero mode can also decay into the SM particles through the mixing with the bounded massive modes due to the modulation.
In the appendix \ref{sec:classical}, we discuss the classical motion of the fermion along the curved string.
In the classical treatment the mixing with the bounded massive modes should correspond to the oscillation of the fermion around the center of the string.
We also note that we have eventually approximated 
the propagator in Eq.\,\eqref{eq:propagator} by $m_\psi^{-1}$.
Since the fermion mass inside the cosmic string is smaller than $m_\psi$, 
the actual decay rate could be enhanced compared to 
the present estimate.
At any rate, the decay rate given in Eq.\,\eqref{eq:decayrate_master} should be regarded as the lower limit of the decay rate.

\subsection{Sinuous Modulation}
As a simple example of the modulation, let us consider a sinuous modulation with 
\begin{align}
    f(z) =  \frac{\varepsilon}{m_\phi}\sin\left(\frac{2\pi z}{L}\right)\ ,
\end{align}
where $\varepsilon\ll 1$ and $L/2\pi \gg m_\phi^{-1}$.
The Fourier coefficients of this modulation are given by 
\begin{align}
&c_1 = i \frac{\varepsilon}{2m_\phi}\ ,\\
&c_n = 0 \ , \quad (n>1) \ ,
\end{align}
where $c_{-n} = c_{n}^*$.
The resultant decay rate is given by
\begin{align}
\label{eq:decayrate_sin}
   \Gamma(E) &\simeq 
\frac{\pi^2}{24}\frac{\varepsilon^2|y_D|^2E}{m_\phi^2L^2}
\left(1-\frac{\pi}{EL}\right)\xi^2\left(\frac{m_\phi}{m_\psi}\right)\ .
\end{align}
The decay rate is non-vanishing only for $E > \pi/L$.

For a sinuous modulation in Eq.\,\eqref{eq:sinuous}, the maximal curvature radius is 
\begin{align}
    R = \frac{L^2}{(2\pi)^2 \varepsilon m_\phi}\ .
\end{align}
In terms of this curvature radius, the decay rate is given by,
\begin{align}
\label{eq:decayrate_sin_R}
   \Gamma(E) &\simeq 
\frac{1}{96}\varepsilon|y_D|^2\frac{E}{m_\phi R}
\xi^2\left(\frac{m_\phi}{m_\psi}\right)\ .
\end{align}
Here, we have assumed $E \gg 1/L$.
This expression suggests that the decay rate in a curved string with a large curvature radius $R$ is suppressed by a single power of $R$.
In the appendix \ref{sec:SingleGaussian}, we calculate the decay rate of the fermion zero mode in a straight string with a Gaussian modulation.
There, we find the same dependence on the curvature radius $R$ 
with Eq.\,\eqref{eq:decayrate_sin_R}.

\subsection{Piecewise Circle Modulation}
\label{sec:pCircle}
\begin{figure}[tbp]
\centering{\includegraphics[width=0.4\textwidth]{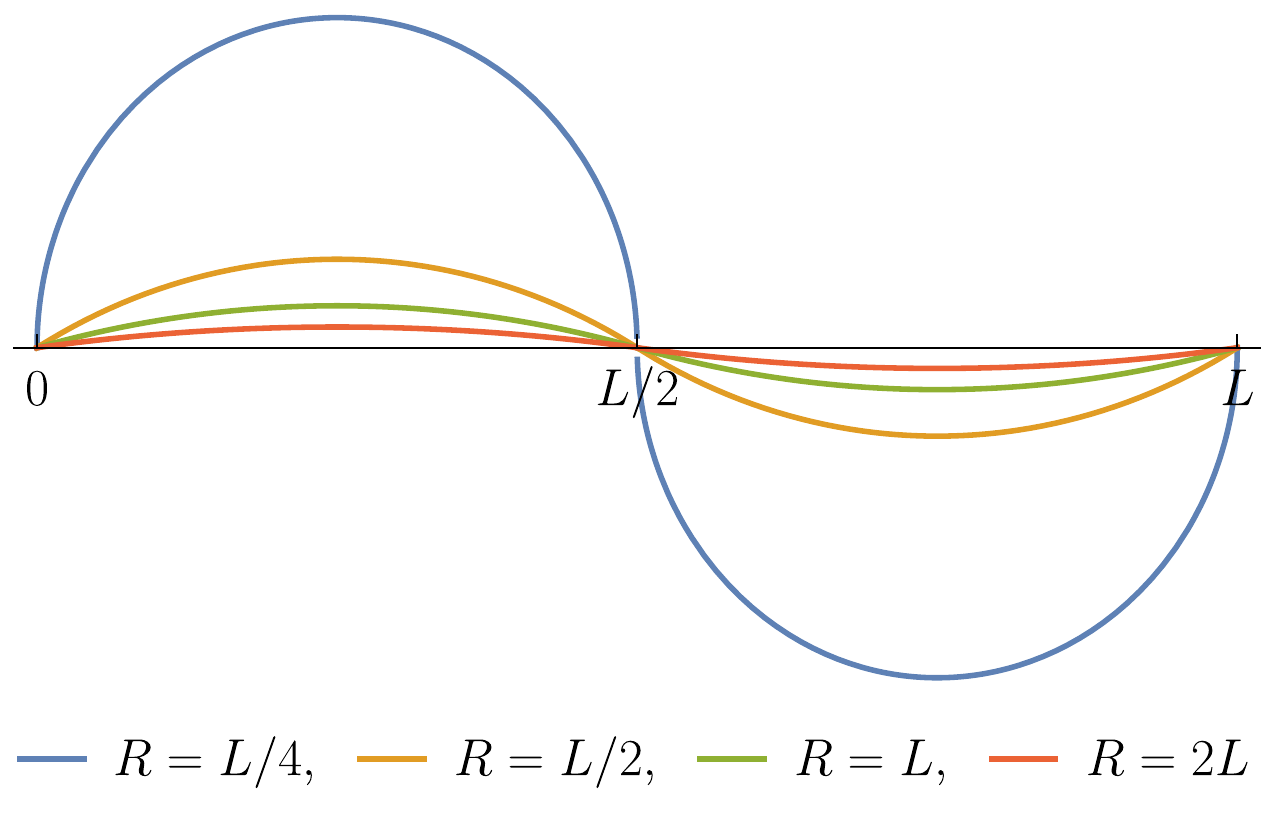}} 
\caption{
The modulation of the periodic piecewise circle with the radius $R$ and the period $L$.
The maximum amplitude of the modulation is given by $f(L/4) \simeq L^2/32R$ for $R\gg L$.
}
\label{fig:pcircle}
\end{figure}
Next, let us consider a periodic modulation given by a piecewise circle with the curvature radius $R$ and a period $L$, i.e., $ f(z+ n L) = f(z)$ with integers $n$, 
\begin{align}
 &f(z) = 
 \begin{cases}
  \sqrt{R^2 - (z-
  L/4)^2} - \sqrt{R^2 - L^2/16}\ , & (0< z < L /2)  \ , \\
   -\sqrt{R^2 - (z-3L/4)^2} + \sqrt{R^2 - L^2/16}\ , & ( L/2< z < L)\ , 
    \end{cases}
 \end{align}
(Fig.\,\ref{fig:pcircle}).
The modulation is piecewise smooth.
The maximum amplitude of the modulation is given by,
\begin{align}
 f(z = L/4) \simeq \frac{L^2}{32 R}\ ,
\end{align}
for $R\gg L$.
Accordingly, the perturbative condition, $|\delta h^2| \ll |\delta h|$, is satisfied for 
\begin{align}
\label{eq:perturbL}
    L \ll \sqrt{\frac{32 R}{m_\phi}} \ll R \ ,
\end{align}
where the final inequality comes from the assumption, $R \gg m_\phi^{-1}$.

In this case, the Fourier coefficients for large $R$ are given by,
\begin{align}
\label{eq:c2n+1}
    c_{2n+1} = \frac{iL^2}{2\pi^3(2n+1)^3R}, \quad c_{2n} = 0\ ,
\end{align}
where $n$ is an integer.
As a result, the decay rate of the fermion zero mode on a piecewise circle is given by,
\begin{align}
\label{eq:decayrate_piecewise}
    \Gamma(E) \simeq 
\frac{|y_D|^2 E}{2^{8}3^2} \frac{L^2}{R^2}\left(1-\frac{1}{EL}\frac{84\zeta(3)}{\pi^3}\right)\xi^2\left(\frac{m_\phi}{m_\psi}\right)\ .
\end{align}

\begin{figure}[tbp]
\centering{\includegraphics[width=0.8\textwidth]{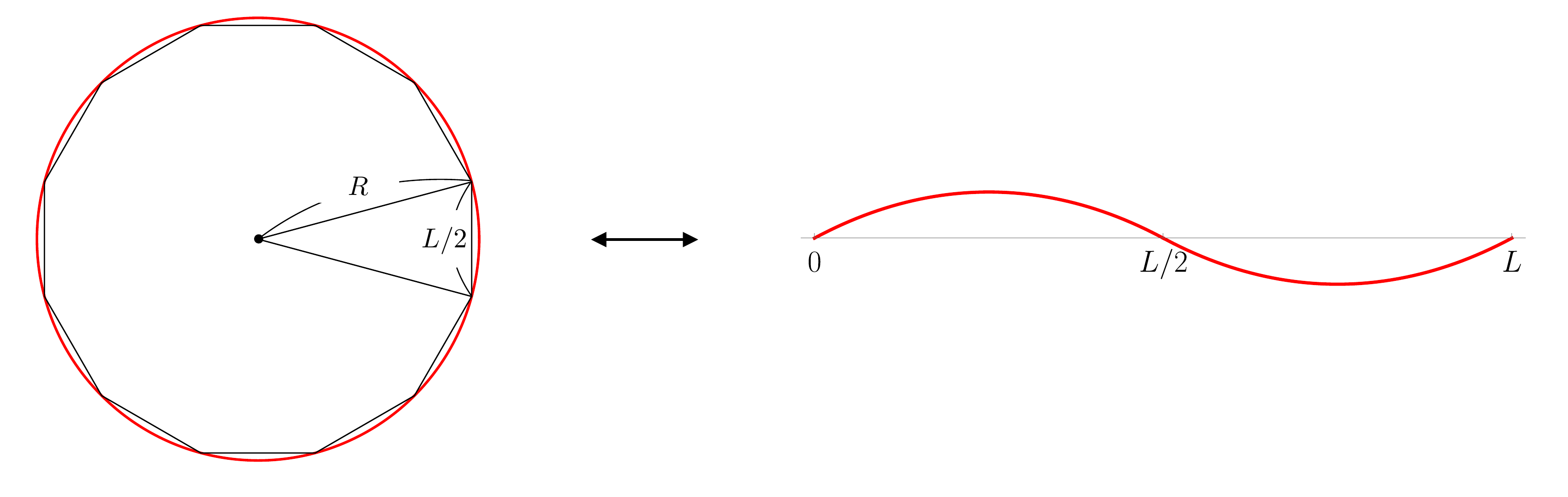}} 
\caption{The approximation of the circular string by taking its segments with a chord length $L/2$. 
For a perturbative analysis, we take a short chord length which satisfies Eq.\,\eqref{eq:perturbL}.
}
\label{fig:segments}
\end{figure}

As an application of the decay rate in Eq.\,\eqref{eq:decayrate_piecewise}, let us consider a closed string loop.
For simplicity, we assume a circular loop with 
a large radius $R$ compared to the string core size $m_\phi^{-1}$.
To apply the perturbative analysis, 
we take segments of the ring with the chord length of $L/2$ (Fig.\,\ref{fig:segments}). Here, $L$ is taken to satisfy the perturbativity condition in Eq.\,\eqref{eq:perturbL}.
Then, the decay rate for a given $L$ can be approximated by Eq.\,\eqref{eq:decayrate_piecewise}.
For the optimal lower limit on the decay rate, we take 
\begin{align}
L\simeq \sqrt{\frac{32 R}{m_\phi}}\ .
\end{align}
Then, we obtain the lower limit on the decay rate of the fermion zero mode on a ring,
\begin{align} \label{eq:ring_decay}
    \Gamma(E) \gtrsim 
     \frac{|y_D|^2E}{72 m_\phi R}
 \xi^2\left(
 \frac{m_\phi}{m_\psi}\right)\ ,
\end{align}
where we have assumed $E \gg \pi/\sqrt{32 R m_{\phi}^{-1}}$.

The above lower limit can be applied to a long string with a curvature radius of $R$.
In the early Universe, the typical macroscopic curvature radius of the string is 
 $\order{H^{-1}}$.
Thus, the lower limit of the decay rate is 
\begin{align} 
    \Gamma(E) \gtrsim 
     \frac{|y_D|^2E}{72 m_\phi}
 \xi^2\left(
 \frac{m_\phi}{m_\psi}\right) \frac{H}{\kappa}\ ,
\end{align}
where we assume $R = \kappa/H$ with $\kappa \lesssim \order{1}$.
Therefore, we find that the high energy modes, 
\begin{align}
    E \gtrsim      \frac{72 m_\phi\kappa}{|y_D|^2}
 \xi^{-2}\left(
 \frac{m_\phi}{m_\psi}\right) \ ,
\end{align}
cannot survive the cosmic time.

\subsection{Thermal Modulation}
Finally, let us roughly estimate the fermion decay rate in the presence of thermal fluctuations at the cosmic temperature $T \ll v_{\mathrm{PQ}}$.
Along the straight string, there are two translational massless moduli fields, $\delta \phi_{x,y}(t,z)$ (see e.g., Ref.\cite{Shifman:2012zz}). 
These moduli fields correspond to the Nambu-Goldstone modes associated with spontaneous breaking of the translational symmetry in the transverse dimension due to the cosmic string.
The moduli fields are thermalized with the SM thermal bath through the coupling to the gluons.
The thermal fluctuation of the moduli fields 
 is roughly given by
 \begin{align}
    \delta \phi_{x,y} \sim T \sin(Tz)  \ ,
 \end{align}
where the typical momentum of the thermal fluctuation is of $\order{T}$.
Since the moduli fields are the modulation of the string configuration, $\delta h$, the above fluctuation corresponds to
\begin{align}
f(z) \sim \frac{T}{m_\phi v_{\mathrm{PQ}}} \sin(Tz)\ ,
\end{align}
where we approximate $dh(\rho)/d\rho \sim m_\phi$.
Here, we neglect the time-dependence of the moduli fluctuation, which does not affect the order of magnitude estimate of the decay rate for $T \ll E$.
By substituting the fluctuation into Eq.\,\eqref{eq:decayrate_sin}, we obtain,
\begin{align}
\label{eq:decayrate_thermal}
    \Gamma(E)\sim \frac{1}{96}
    \frac{|y_D|^2 T^4 E}{m_\phi^2 v_{\mathrm{PQ}}^2}
    \xi^2\left(\frac{m_\phi}{m_\psi}\right) \ .
\end{align}
Thus, the decay of the fermion zero mode caused by thermal fluctuation becomes smaller than the Hubble expansion rate at a temperature lower than,
\begin{align}
T \lesssim 20\frac{m_\phi}{|y_D|\xi }
\left(\frac{v_{\mathrm{PQ}}}{E}\right)^{1/2}\left(\frac{v_{\mathrm{PQ}}}{M_\mathrm{Pl}}\right)^{1/2}
\sim 4\times 10^5\,\mathrm{GeV}\times
\frac{1}{|y_D|\xi}
\left(\frac{m_\phi}{10^9\,\mathrm{GeV}}\right)
\left(\frac{v_{\mathrm{PQ}}}{10^9{\,\mathrm{GeV}}}\right)^{1/2} 
\left(\frac{v_{\mathrm{PQ}}}{E}\right)^{1/2}
\ . 
\end{align}
Here, $M_\mathrm{Pl}$ is the reduced Planck scale.
Below this temperature, the decay of the fermion zero mode induced 
by the thermal fluctuation becomes ineffective.

\section{Fate of Vorton}
\label{sec:vorton}
\subsection{Vorton Radius}
As discussed in Ref.~\cite{Fukuda:2020kym}, the superconductivity of the axion string may have significant effects on the cosmological evolution of the axion string network. 
In particular, the formation of the stable configuration, the Vorton, may contribute to the dark matter density.
We review how the Vorton is stabilized by considering a string loop with a length $L_\mathrm{loop}$ and a total QED charge, $Q$, carried by the fermion zero modes.
Here, we assume that the length of the string loop is much longer than its core size of $\mathcal{O}(m_\phi^{-1})$.
If there is no charge dissipation nor leakage, $Q$ of the string loop is conserved.
Due to the Fermi statistics, the total charge is related to the Fermi momentum of the fermion zero modes, $\varepsilon_F$, via,
\begin{align}
    \label{eq:Q}
    Q = \frac{1}{2\pi} N_c q_\psi\varepsilon_F L_\mathrm{loop} \ ,
\end{align}
where $N_c = 3$ is the color factor.
Here, we approximate the distribution of the fermion zero modes by the Fermi-Dirac distribution at the zero temperature.
The total energy of the fermion zero modes trapped in the string loop is
\begin{align}
    E_Q = \frac{1}{4\pi} N_c \varepsilon_F^2 L_\mathrm{loop} = 
    \frac{\pi Q^2}{N_cq_\psi^2L_\mathrm{loop}}\ .
\end{align}
The total energy of the string is given by the sum of the fermion energy and the weight of the string loop,
\begin{align}
    E(L_\mathrm{loop}) \sim \mu L_\mathrm{loop} + \frac{\pi Q^2}{N_cq_\psi^2L_\mathrm{loop}}\ ,
\end{align}
with $\mu$ being the string tension,
\begin{align}
    \mu \simeq 2\pi v_{\mathrm{PQ}}^2 \log(m_{\phi} L_\mathrm{loop}) \ .
\end{align}

The string loop shrinks by emitting the axions.
As $L_\mathrm{loop}$ decreases, the Fermi momentum increases 
due to the charge conservation.
When $L_\mathrm{loop}$ decreases,
the kinetic energy of the fermion zero modes and the string tension balances and 
the string loop is stabilized for 
\begin{align}
\label{eq:vortonlength}
    L_\mathrm{loop} \sim \sqrt{\frac{1}{2N_c \log(m_{\phi} L_\mathrm{loop})}}\frac{Q}{q_\psi v_{\mathrm{PQ}}}\ .
\end{align}
The corresponding stable configuration is called the Vorton.
Notably, the Fermi momentum in the Vorton does not depend on the total charge, and is given by,
\begin{align}
\varepsilon_F \sim 2\pi \sqrt{\frac{2\log(m_{\phi}L_\mathrm{loop})}{N_c}}v_{\mathrm{PQ}} > v_{\mathrm{PQ}}\ .
\label{eq:FermiMomentum}
\end{align}

In this paper, we do not discuss the formation and the evolution of the Vortons in detail.
Instead, we consider a Vorton with a typical total charge formed at the cosmic temperature $T_\mathrm{form}$ with a loop length $\order{H^{-1}}$,
\begin{align}
 Q 
 \sim\left(\frac{M_\mathrm{Pl}}{v_\mathrm{PQ}}\right)^{1/2}\ .
\end{align}
according to Ref.\,\cite{Fukuda:2020kym}.
Here we take the $T \sim m_\phi \sim v_\mathrm{PQ}$.
If there is no charge leakage, the radius of the Vorton is expected to be, 
\begin{align}
\label{eq:radius}
    R \sim \frac{Q}{v_{\mathrm{PQ}}} \sim \frac{1}{v_{\mathrm{PQ}}}\left(\frac{M_\mathrm{Pl}}{v_\mathrm{PQ}}\right)^{1/2}\ .
\end{align}
In the following argument, we discuss the fate of the Vorton with the radius of this order.

\subsection{Decay of Fermion Zero Mode in Vorton}
As we have discussed in the previous section, the fermion zero mode decays through the decay operator when the string is curved.
As we have also seen in Sec.\,\ref{sec:pCircle}, the perturbative analysis can be used to give a lower limit on the decay rate of string loop with a curvature radius, $R \gg m_\phi^{-1}$.

Now, let us apply the lower limit on the decay rate in Eq.\,\eqref{eq:ring_decay} to discuss the stability of the Vorton.
By substituting the Vorton radius in Eq.\,\eqref{eq:radius},
the lower limit on the decay rate is,
\begin{align}
\label{eq:decayrate_vorton}
    \Gamma(E)\gtrsim  \frac{|y_D|^2  v_\mathrm{PQ}}{72m_\phi}
    \left(\frac{v_\mathrm{PQ}}{M_\mathrm{Pl}}\right)^{1/2}
    E \,\xi^2\left(\frac{m_\phi}{m_\psi}\right) \ .
\end{align}
This decay rate is much larger than the Hubble expansion
rate, $H\simeq T^2/M_\mathrm{Pl}$ $(T\lesssim v_{\mathrm{PQ}})$, especially for the 
fermion zero mode with the Fermi momentum, $E = \varepsilon_F$, unless $y_D$ is highly suppressed for $T < v_\mathrm{PQ}$.
Thus, we find that the high momentum fermion zero mode is short lived.

Once the high momentum zero modes decay, the total charge decreases and its radius shrinks.%
\footnote{The circular string loop starts spinning when it emits the SM particles by the decay of the fermion zero mode. However, the string loop shrinks by emitting the axions
even if it is spinning
since the axions carry away the angular momentum of the spinning string loop.}
Once it shrinks, then,  remaining fermion zero modes on the string obtain high momenta
and the Fermi momentum is again given by Eq.\,\eqref{eq:FermiMomentum}.
Thus, the lower limit on the rate of the charge leakage is given by,
\begin{align}
    -\frac{\dot{Q}}{Q} \gtrsim
  \frac{|y_D|^2 v_\mathrm{PQ}}{144 m_\phi} \left(\frac{v_\mathrm{PQ}}{M_\mathrm{Pl}}\right)^{1/2} 
    \varepsilon_F \xi^2\left(\frac{m_\phi}{m_\psi}\right)
    = \frac{\pi  v_\mathrm{PQ}\varepsilon_F}{9 m_\phi m_\psi} \left(\frac{v_\mathrm{PQ}}{M_\mathrm{Pl}}\right)^{1/2}  \xi^2\left(\frac{m_\phi}{m_\psi}\right)\times \Gamma_D\ .
\end{align}
The lifetime of the Vorton for $m_\phi \gg m_\psi$ is
\begin{align}
    \tau_{\rm Vorton} \sim  -\left(\frac{\dot{Q}}{Q}\right)^{-1} & \lesssim  
    10^6 \times  \tau_{D} \left( \frac{v_{\mathrm{PQ}}}{10^9~\mathrm{GeV}}\right)^{-1/2} \left(\frac{m_\phi^3}{m_{\psi}v_{\mathrm{PQ}}^2}\right) \\
    & \simeq |y_D|^{-2} \times 10^{-26}~\mathrm{sec} \left( \frac{v_{\mathrm{PQ}}}{10^9~\mathrm{GeV}}\right)^{-3/2}
    \left(\frac{m_\phi^3}{m_{\psi}^2v_{\mathrm{PQ}}}\right) \ ,
\end{align}
where we have used $\varepsilon_F \sim v_\mathrm{PQ}$.
Therefore, we expect that the Vorton immediately disappears by losing its charge through the decay operator in Eq.\,\eqref{eq:decay}.

One caveat is that the Vorton
can have magnetic fields up to of
$\order{\varepsilon_F /4\pi^2\rho_*}$,
where $\rho_* =
\order{\max[{m_\phi^{-1}},{m_\psi^{-1}}]}$ represents
a typical thickness of the 
fermion zero mode current.
Since the magnetic fields affect the wave functions of the charged particles,
they can also affect the decay rate.
For the decay
caused by higher 
Fourier modes in Eq.\,\eqref{eq:c2n+1}, however, 
the plane wave approximation
for the final state
is valid.
The decay rate with 
such higher Fourier modes
is high enough as seen in 
Eq.\,\eqref{eq:c2n+1}.
Thus, we do not expect that the magnetic field makes the Vorton have a 
cosmological lifetime.

\section{Conclusion}
In this paper, we studied the stability of the fermion which carries the superconducting current in the axion string. 
In particular, we discussed the effect of the decay operator of the charge carriers.
We found the superconductivity is indeed stable for a straight string or infinitely small string core size.
We also found that the charge carriers decay when the string with a finite core size is curved.
We obtained the lower limit on the decay rate 
which is suppressed only by a power law of the curvature radius of the string. 
As a result, the charge carriers of the Vorton are not long-lived, and hence, the Vortons that appear in the axion models do not have a lifetime of the cosmological timescale.
Our analysis on the decay rate of the fermion zero mode can be applied for generic fermionic superconductive strings.

Although we have argued that the Vorton does not contribute to the dark matter density, the superconductive nature of the global string in the axion model may affect the cosmology in the axion model.
For example, Ref.~\cite{Fukuda:2020kym} 
pointed out that the energy density of the cosmic string can deviate from the scaling law if it exhibits the superconductivity.
Therefore, further study of the effects of the superconductivity of the axion string is required.

Finally, let us comment on the possibility 
of the stable Vorton.
If the fermions are stable in the vacuum, 
the corresponding fermion zero modes in the string are stable.
Thus, for example, if the electron zero mode 
appears in the string and the charge 
of the Vorton is large enough,
the Vorton can be stable 
(see Ref.\,\cite{Witten:1984eb} for the $\mathrm{O}(10)$ model
and see also Ref.\,\cite{Abe:2020ure} in the context of the DFSZ axion model~\cite{Zhitnitsky:1980tq,Dine:1981rt}).
It is, however, not clear how such a large charge is achieved in the cosmological evolution
for the electron zero modes. 
We leave the details of these possibilities for future work.

%%%%%%%%%%%%%%%%%%%%%%%%%%%%%%%%%%%%
\section*{Acknowledgments}
%%%%%%%%%%%%%%%%%%%%%%%%%%%%%%%%%%%%
 The authors thank H.~Fukuda for useful comments.
This work is supported by Grant-in-Aid for Scientific Research from the Ministry of Education, Culture, Sports, Science, and Technology (MEXT), Japan, 17H02878 (M.I. and S.S.), 18H05542 (M.I.), 18K13535, 19H04609, 20H01895 and 20H05860 (S.S.), and by World Premier International Research Center Initiative (WPI), MEXT, Japan. 
This work is also supported by the Advanced Leading Graduate Course for Photon Science (S.K.), the JSPS Research Fellowships for Young Scientists (S.K.) and International Graduate Program for Excellence in Earth-Space Science (Y.N.).

\appendix
\section{Classical Motion of Particle along Curved String}
\label{sec:classical}

In this appendix, we discuss the classical motion of the fermion zero mode along the curved string.
Let us consider the case that the center of the string is in two-dimensional space,
\begin{align}
    (x,y,z)=(0,f(z),z)\ ,
\end{align}
where the configuration of the PQ field
is given by $h_{f}(\vec{r})$ around the center.
We treat the fermion zero mode as a classical particle whose Hamiltonian 
is given by,
\begin{align}
    H_\mathrm{cl} = \sqrt{p_y^2 + p_z^2 + m_\psi^2 h_{f}^2(y,z)} \ ,
\end{align}
where $h_f = 0 $ at the center of the string.
When the particle travels through the center of the string, it behaves as a massless particle.

\begin{figure}
    \centering
    \includegraphics[width=0.6\linewidth]{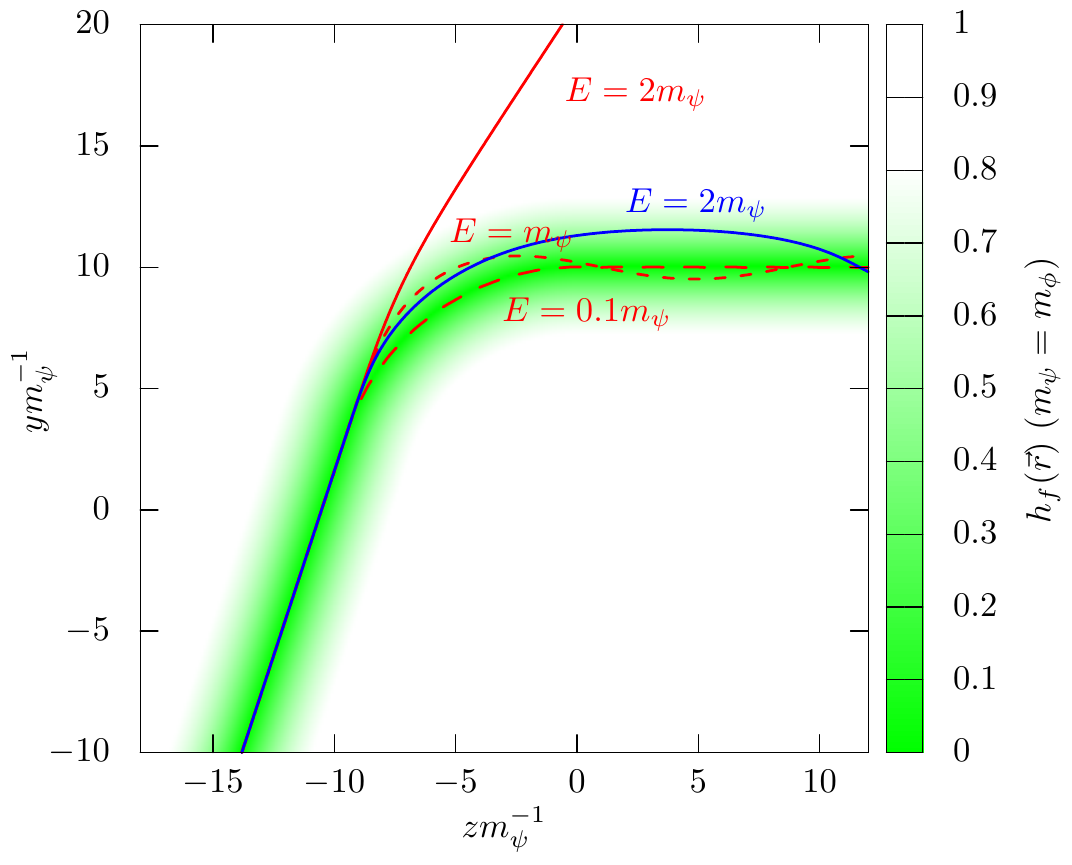}
    \caption{
    Examples of the classical trajectories of the particle.
    Here we take $m_{\phi} =  m_{\psi}$ for the red lines and $m_{\phi} =  3 m_{\psi}$ for the blue line.
    }
    \label{fig:classical_example}
\end{figure}

The equations of motion are given by
\begin{align}
    \dot{\vec{r}} &= \frac{\vec{p}}{\sqrt{|\vec{p}|^2 + m_\psi^2 h_f^2}},\\
    \dot{\vec{p}} &= -\frac{ m_{\psi}^2  h_f \vec{\nabla} h}{\sqrt{|\vec{p}|^2 + m_\psi^2 h_f^2}}.
\end{align}
In Fig.~\ref{fig:classical_example}, we show examples of the classical trajectories of the particle.
We approximate $h_f(y,z) = h(\mathrm{distance~between~}(y,z)\mathrm{~and~} f(z))$, where $h(\rho)$ is the profile of the straight string  in Eq.\,\eqref{eq:h}.
As we can see from Fig.~\ref{fig:classical_example}, if the energy of the particle is high and the string curve is sharp, the particle can escape from the string.
We also see that the smaller string core size prevents the particle from escaping, even if the particle's momentum exceeds the free particle mass.
As discussed in Sec.\,\ref{sec:classical_escape}, 
once the particle escapes from the string, it behaves as a free massive particle.
Thus, in the presence of the decay operator, it decays immediately through the decay operator.

Even if the particle is not energetic enough to jump out the string, the classical trajectory of the particle is off from the center of the string core and oscillates about the string center, once the particle passes the curve.
During the oscillation, particle gets a non-zero mass thorough the VEV of the PQ field as $m_{\psi,\mathrm{eff}}  = m_{\psi} h_f(\vec r)$.
Once the particle obtains the effective mass, the particle will decay into the lighter SM particles through the decay operator\,\eqref{eq:decay}.
To estimate the effective decay rate in-flight, we define:
\begin{align}
\Gamma_\mathrm{eff}(E) = \frac{|y_D|^2}{16\pi}
\frac{1}{T}\int_0^T dt
\frac{m_\psi^2 h_f^2(y(t),z(t))}{H_{\mathrm{cl}}}\ ,
\end{align}
where the factor of $m_\psi h_f/H_{\mathrm{cl}}$ is the suppression due to the Lorentz boost.

\begin{figure}
    \centering
    \includegraphics[width=0.5\linewidth]{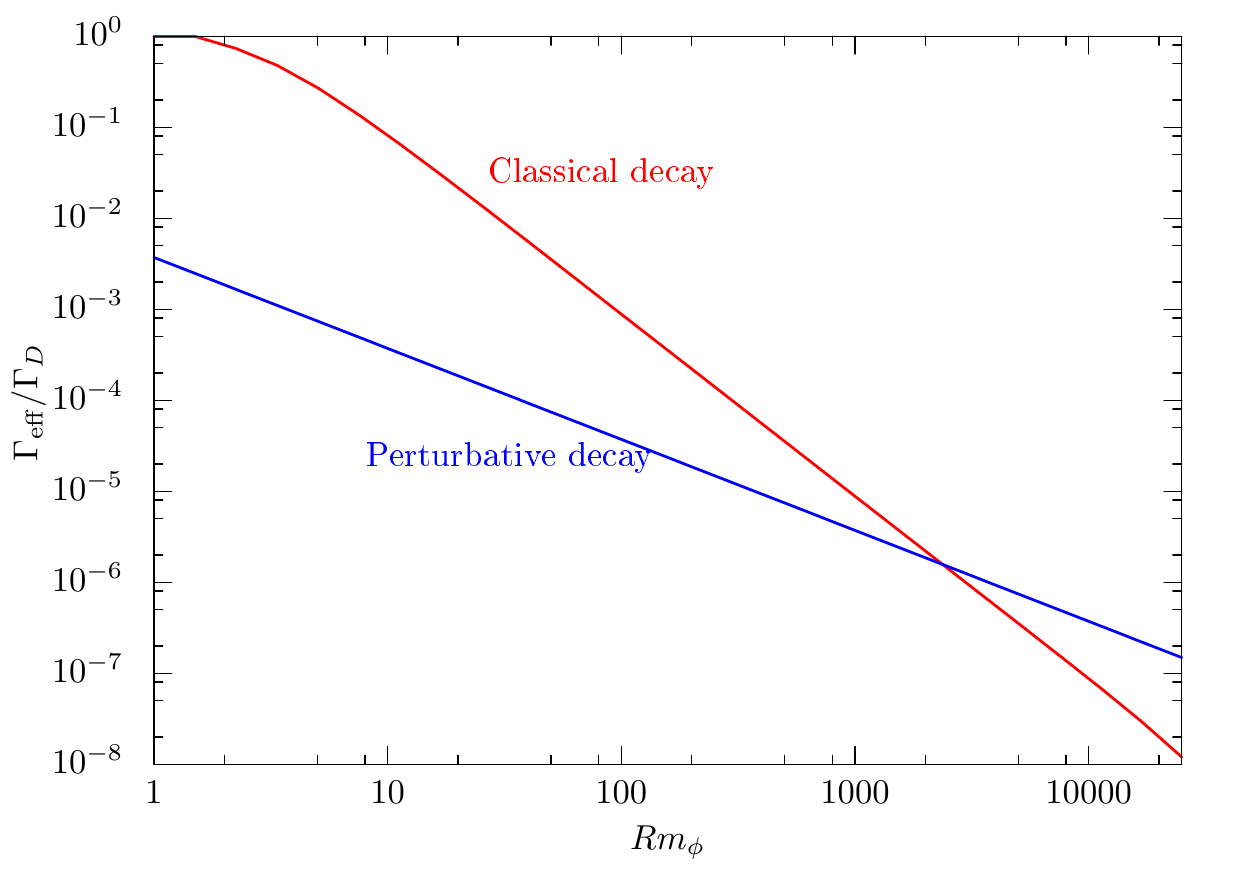}
    \caption{
    The ratio of the effective decay rate of the fermion on the string ring to that of the free fermion $\Gamma_D$ in Eq.\,\eqref{eq:free_decay}. 
    The center of the ring is at the origin and its radius is $R$.
    Here we set $m_{\psi} = m_{\phi}$ and the initial conditions of the particle are $\vec{r} = (R,0)$ and $\vec{p} = (0,m_{\phi})$. 
    We approximate $h_f(\vec{r}) = h( \abs{ \abs{\vec{r}} -R }) $ where 
    $h$ is the profile in Eq.\,\eqref{eq:h}.
    For a comparison, we also show the estimate of the lower limit on the decay rate obtained by the perturbative method \eqref{eq:ring_decay} with the blue line. 
    }
    \label{fig:classical_ring}
\end{figure}

For demonstration, we consider the effective decay rate of a particle trapped in a ring of the  string in Fig.\,\ref{fig:classical_ring}.
The figure shows that the perturbative analysis in Sec.\,\ref{sec:decay_master} predicts a larger decay rate 
than the classical estimation for a large ring radius.
In the perturbative analysis, 
we consider the decay of the fermion 
through the virtual free massive modes.
The present classical treatment, on the other hand, corresponds to the mixing between the fermion zero mode 
and the massive bounded mode due to the string curve.
Such effects are neglected in the perturbative
analysis.
Therefore, the decay processes in the perturbative analysis and in 
the classical treatment are independent.
In both cases, the decay rates are suppressed for a large ring radius.
Despite the suppression, 
the either decay rate is too large to keep the Vorton stable.
Note that the classical treatment is not valid for $m_\psi \ll m_\phi$,
where the quantum mechanical treatment is more important.

\section{Phase Space Integration}
\label{sec:phasespace}
In this appendix, we calculate the phase space integration in Eq.\,\eqref{eq:decayrate_integral}.
In the polar coordinate,
\begin{align}
&d^3 \mathbf{p}_q =|\mathbf{p}_q|p_q^0  dp_q^0d\cos\theta_q d\varphi_q\ , \\
&d^3 \mathbf{p}_H =|\mathbf{p}_H|p_H^0  dp_H^0d\cos\theta_H d\varphi_H\ .
\end{align}
By defining,
\begin{align}
    &p^0_q = \frac{1}{2}(E_f + E_-)\ ,\\
    &p^0_H = \frac{1}{2}(E_f - E_-)\ ,
\end{align}
 and using 
\begin{align}
    dp^0_q dp^0_H = \frac{1}{2}dE_f dE_- ,
\end{align}
the integration over $E_f$ gives $E_f = E$ due to the delta function in Eq.\,\eqref{eq:decayrate_integral}.
By neglecting the masses of the quark and the Higgs doublets, 
the integration over $E_-$ also gives,
\begin{align}
E_-  = \frac{(2-\cos\theta_H-\cos\theta_q)E-2k_n}{\cos\theta_q-\cos\theta_H}\ .
\end{align}
The kinematical constraint on $E_-$
is given by $-E < E_- < E$, which is satisfied for a range of the angles,
\begin{align}
    \left(-1 <\cos\theta_H < 1 - \frac{k_n}{E}\right) \land 
    \left(1 - \frac{k_n}{E} <\cos\theta_q < 1\right)\ ,
\end{align}
and 
\begin{align}
    \left(1- \frac{k_n}{E}  <\cos\theta_H < 1\right) \land 
    \left(-1  <\cos\theta_q < 1-\frac{k_n}{E}\right)\ .
\end{align}
Here, $k_n < 2 E$.
As a result, we find
\begin{align}
    &\int\!\!\! \frac{d^3\mathbf{p}_q}{(2\pi)^3 2p_q^0}
\frac{d^3\mathbf{p}_H}{(2\pi)^3 2p_H^0}
(2\pi)^4\delta(E-E_f)\delta(E-p_{f}^z-k_n) (p_q^0 - p_q^3)\\
&= \frac{1}{4}\int d\cos\theta_H d \cos\theta_q |\mathbf{p}_H|
|\mathbf{p}_q|^2(1-\cos\theta_q)(\cos\theta_H-\cos\theta_q)^{-1}\ , \\
&=
\frac{1}{24} E^3 \mathcal{F}(k_n/E)\ , 
\end{align}
where 
\begin{align}
    \mathcal{F}(x) = x^2(1-x/2)\ .
\end{align}

\section{Gaussian Modulation}
\label{sec:SingleGaussian}
In Sec.~\ref{sec:decayonmodulation}, we considered a periodic modulation.
In this section, we consider the decay of the fermion zero mode in a string with a single Gaussian modulation,
\begin{align}
\label{eq:gaussian}
    f(z) = \frac{\varepsilon}{m_\phi} e^{-\frac{z^2}{2\sigma_z^2 }} \ .
\end{align}
In this case, the matrix element becomes,
\begin{align}
T &\simeq 
i(2\pi)\delta(E-E_f)
\frac{y_D^*\varepsilon}{m_\phi} \bar{u}_{\mathrm{q}}P_R
\int d^3x e^{i\varphi} \sin\varphi 
\frac{dh}{d\rho}u(\rho)e^{-\frac{z^2}{2\sigma_z^2}}e^{iEz}e^{-i\mathbf{p}_f\cdot\mathbf{x}} \ , \\ 
&\simeq (2\pi)^{5/2}\delta(E-E_f)
\sigma_z e^{-\frac{\sigma_z^2}{2}(E-p_f^3)^2}\frac{y_D^*\varepsilon}{m_\phi} \bar{u}_{\mathrm{q}}P_R\xi
\left(\frac{m_\phi}{m_\psi}\right)\eta\ , 
\end{align}
where we have used
the same approximation in Sec.\,\ref{sec:decay_master}.
By dividing the squared amplitude by the total time,  the decay rate of the fermion zero mode is given by,
\begin{align}
\label{eq:decayrate_gauss_integral}
\Gamma(E) \times L_{\mathrm{str}}\simeq 
\frac{|y_D|^2 \varepsilon^2}{m_\phi^2} \xi^2
\left(\frac{m_\phi}{m_\psi}\right)\sigma_z^2\int\!\!\! \frac{d^3\mathbf{p}_q}{(2\pi)^3 2p_q^0}
\frac{d^3\mathbf{p}_H}{(2\pi)^3 2p_H^0}
(2\pi)^4\delta(E-E_f) (p_q^0 - p_q^z)e^{-\frac{\sigma_z^2}{2}(E-p_f^z)^2}\ ,
\end{align}
where $L_\mathrm{str} = (2\pi)\delta(p_z)|_{p_z = 0}$.

We take the state normalized so that the number of the fermion zero modes is constant for a unit length of the string (Eq.\,\eqref{eq:normalization}).
Thus, the decay rate on average is vanishing for $L_{\mathrm{str}}\to \infty$, since most of the fermion zero mode is far away from the modulation on the string.
However, we are interested in the decay rate of the fermion zero mode when it is passing through the modulation.
Thus, instead of the decay rate on average,  we should consider the effective decay rate which is
given by dividing $\Gamma(E)$ by
the probability to find the Gaussian modulation, $P_m = \order{\sigma_z/L_{\mathrm{str}}}$.
As a result, we find
\begin{align}
\label{eq:decayrate_gauss}
\Gamma_\mathrm{eff}(E)&\sim 
\frac{|y_D|^2 \varepsilon^2}{m_\phi^2} \xi^2
\left(\frac{m_\phi}{m_\psi}\right)\sigma_z\int\!\!\! \frac{d^3\mathbf{p}_q}{(2\pi)^3 2p_q^0}
\frac{d^3\mathbf{p}_H}{(2\pi)^3 2p_H^0}
(2\pi)^4\delta(E-E_f) (p_q^0 - p_q^3)e^{-\frac{\sigma_z^2}{2}(E-p_f^z)^2}\ ,\\
 &= 
\frac{|y_D|^2 \varepsilon^2}{8m_\phi^2} \xi^2
\left(\frac{m_\phi}{m_\psi}\right)\sigma_z\int\!
 dE_-d\cos\theta_q d\cos\theta_H
 |\mathbf{p}_H||\mathbf{p}_q|^2
 (1 - \cos\theta_q)\notag\\
 & \hspace{4cm}\times\exp\left[-\frac{\sigma_z^2}{2}
  (E-(|\mathbf{p}_q|\cos\theta_q + |\mathbf{p}_H|\cos\theta_H))^2\right]
 \ ,\\
 &=\frac{|y_D|^2 \varepsilon^2 }{8m_\phi^2}\xi^2
\left(\frac{m_\phi}{m_\psi}\right) \sigma_z E^4 \mathcal{G}(\sigma_z E)\ ,
\end{align}
where the integration is given by,
\begin{align}
\mathcal{G}(x) = \frac{-2 + 2 e^{-2 x^2} + \sqrt{2 \pi} x \erf(\sqrt{2} x)}{2 x^4}  \simeq \frac{1.25}{x^3} \left(1-\frac{4}{5 x}\right)\ . 
\end{align}
Thus, for $\sigma_z E \gg 1$, we find that 
the effective decay rate is approximately given by,
\begin{align}
    \Gamma_{\mathrm{eff}}(E)\sim\frac{1}{6}\frac{|y_D|^2 \varepsilon^2 }{m_\phi^2} \frac{E}{\sigma_z^2}\xi^2
\left(\frac{m_\phi}{m_\psi}\right)\ .
\end{align}

The maximum curvature radius of the modulation in Eq.\,\eqref{eq:gaussian}
is,
\begin{align}
R = \varepsilon^{-1} m_\phi \sigma_z^2 \ .
\end{align}
Thus, in terms of the curvature radius, the effective decay rate scales as,
\begin{align}
\Gamma_{\mathrm{eff}}(E) \sim\frac{1}{6}\varepsilon|y_D|^2  \frac{E}{m_\phi R}\xi^2
\left(\frac{m_\phi}{m_\psi}\right)\ .
\end{align}
This reproduces the decay rate for a periodic modulation with a curvature radius $R$ in Eq.\,\eqref{eq:decayrate_sin_R}.
This analysis justifies the use of the periodic modulation in which we neglect the damping of $\delta h$ at $|z|\to \infty$.

%%%%%%%%%%%%% References %%%%%%%%%%%%%%%%%%%
\bibliographystyle{apsrev4-1}
\bibliography{ref}

%merlin.mbs apsrev4-1.bst 2010-07-25 4.21a (PWD, AO, DPC) hacked
%Control: key (0)
%Control: author (72) initials jnrlst
%Control: editor formatted (1) identically to author
%Control: production of article title (-1) disabled
%Control: page (0) single
%Control: year (1) truncated
%Control: production of eprint (0) enabled
\begin{thebibliography}{40}%
\makeatletter
\providecommand \@ifxundefined [1]{%
 \@ifx{#1\undefined}
}%
\providecommand \@ifnum [1]{%
 \ifnum #1\expandafter \@firstoftwo
 \else \expandafter \@secondoftwo
 \fi
}%
\providecommand \@ifx [1]{%
 \ifx #1\expandafter \@firstoftwo
 \else \expandafter \@secondoftwo
 \fi
}%
\providecommand \natexlab [1]{#1}%
\providecommand \enquote  [1]{``#1''}%
\providecommand \bibnamefont  [1]{#1}%
\providecommand \bibfnamefont [1]{#1}%
\providecommand \citenamefont [1]{#1}%
\providecommand \href@noop [0]{\@secondoftwo}%
\providecommand \href [0]{\begingroup \@sanitize@url \@href}%
\providecommand \@href[1]{\@@startlink{#1}\@@href}%
\providecommand \@@href[1]{\endgroup#1\@@endlink}%
\providecommand \@sanitize@url [0]{\catcode `\\12\catcode `\$12\catcode
  `\&12\catcode `\#12\catcode `\^12\catcode `\_12\catcode `\%12\relax}%
\providecommand \@@startlink[1]{}%
\providecommand \@@endlink[0]{}%
\providecommand \url  [0]{\begingroup\@sanitize@url \@url }%
\providecommand \@url [1]{\endgroup\@href {#1}{\urlprefix }}%
\providecommand \urlprefix  [0]{URL }%
\providecommand \Eprint [0]{\href }%
\providecommand \doibase [0]{http://dx.doi.org/}%
\providecommand \selectlanguage [0]{\@gobble}%
\providecommand \bibinfo  [0]{\@secondoftwo}%
\providecommand \bibfield  [0]{\@secondoftwo}%
\providecommand \translation [1]{[#1]}%
\providecommand \BibitemOpen [0]{}%
\providecommand \bibitemStop [0]{}%
\providecommand \bibitemNoStop [0]{.\EOS\space}%
\providecommand \EOS [0]{\spacefactor3000\relax}%
\providecommand \BibitemShut  [1]{\csname bibitem#1\endcsname}%
\let\auto@bib@innerbib\@empty
%</preamble>
\bibitem [{\citenamefont {Vilenkin}\ and\ \citenamefont
  {Shellard}(2000)}]{Vilenkin:2000jqa}%
  \BibitemOpen
  \bibfield  {author} {\bibinfo {author} {\bibfnamefont {A.}~\bibnamefont
  {Vilenkin}}\ and\ \bibinfo {author} {\bibfnamefont {E.~S.}\ \bibnamefont
  {Shellard}},\ }\href@noop {} {\emph {\bibinfo {title} {{Cosmic Strings and
  Other Topological Defects}}}}\ (\bibinfo  {publisher} {Cambridge University
  Press},\ \bibinfo {year} {2000})\BibitemShut {NoStop}%
\bibitem [{\citenamefont {Peccei}\ and\ \citenamefont
  {Quinn}(1977)}]{Peccei:1977hh}%
  \BibitemOpen
  \bibfield  {author} {\bibinfo {author} {\bibfnamefont {R.}~\bibnamefont
  {Peccei}}\ and\ \bibinfo {author} {\bibfnamefont {H.~R.}\ \bibnamefont
  {Quinn}},\ }\href {\doibase 10.1103/PhysRevLett.38.1440} {\bibfield
  {journal} {\bibinfo  {journal} {Phys. Rev. Lett.}\ }\textbf {\bibinfo
  {volume} {38}},\ \bibinfo {pages} {1440} (\bibinfo {year}
  {1977})}\BibitemShut {NoStop}%
\bibitem [{\citenamefont {Weinberg}(1978)}]{Weinberg:1977ma}%
  \BibitemOpen
  \bibfield  {author} {\bibinfo {author} {\bibfnamefont {S.}~\bibnamefont
  {Weinberg}},\ }\href {\doibase 10.1103/PhysRevLett.40.223} {\bibfield
  {journal} {\bibinfo  {journal} {Phys. Rev. Lett.}\ }\textbf {\bibinfo
  {volume} {40}},\ \bibinfo {pages} {223} (\bibinfo {year} {1978})}\BibitemShut
  {NoStop}%
\bibitem [{\citenamefont {Wilczek}(1978)}]{Wilczek:1977pj}%
  \BibitemOpen
  \bibfield  {author} {\bibinfo {author} {\bibfnamefont {F.}~\bibnamefont
  {Wilczek}},\ }\href {\doibase 10.1103/PhysRevLett.40.279} {\bibfield
  {journal} {\bibinfo  {journal} {Phys. Rev. Lett.}\ }\textbf {\bibinfo
  {volume} {40}},\ \bibinfo {pages} {279} (\bibinfo {year} {1978})}\BibitemShut
  {NoStop}%
\bibitem [{\citenamefont {Witten}(1985)}]{Witten:1984eb}%
  \BibitemOpen
  \bibfield  {author} {\bibinfo {author} {\bibfnamefont {E.}~\bibnamefont
  {Witten}},\ }\href {\doibase 10.1016/0550-3213(85)90022-7} {\bibfield
  {journal} {\bibinfo  {journal} {Nucl. Phys. B}\ }\textbf {\bibinfo {volume}
  {249}},\ \bibinfo {pages} {557} (\bibinfo {year} {1985})}\BibitemShut
  {NoStop}%
\bibitem [{\citenamefont {Fukuda}\ \emph {et~al.}(2020)\citenamefont {Fukuda},
  \citenamefont {Manohar}, \citenamefont {Murayama},\ and\ \citenamefont
  {Telem}}]{Fukuda:2020kym}%
  \BibitemOpen
  \bibfield  {author} {\bibinfo {author} {\bibfnamefont {H.}~\bibnamefont
  {Fukuda}}, \bibinfo {author} {\bibfnamefont {A.~V.}\ \bibnamefont {Manohar}},
  \bibinfo {author} {\bibfnamefont {H.}~\bibnamefont {Murayama}}, \ and\
  \bibinfo {author} {\bibfnamefont {O.}~\bibnamefont {Telem}},\ }\href@noop {}
  {\  (\bibinfo {year} {2020})},\ \Eprint {http://arxiv.org/abs/2010.02763}
  {arXiv:2010.02763 [hep-ph]} \BibitemShut {NoStop}%
\bibitem [{\citenamefont {Abe}\ \emph {et~al.}(2020)\citenamefont {Abe},
  \citenamefont {Hamada},\ and\ \citenamefont {Yoshioka}}]{Abe:2020ure}%
  \BibitemOpen
  \bibfield  {author} {\bibinfo {author} {\bibfnamefont {Y.}~\bibnamefont
  {Abe}}, \bibinfo {author} {\bibfnamefont {Y.}~\bibnamefont {Hamada}}, \ and\
  \bibinfo {author} {\bibfnamefont {K.}~\bibnamefont {Yoshioka}},\ }\href@noop
  {} {\  (\bibinfo {year} {2020})},\ \Eprint {http://arxiv.org/abs/2010.02834}
  {arXiv:2010.02834 [hep-ph]} \BibitemShut {NoStop}%
\bibitem [{\citenamefont {Agrawal}\ \emph {et~al.}(2020)\citenamefont
  {Agrawal}, \citenamefont {Hook}, \citenamefont {Huang},\ and\ \citenamefont
  {Marques-Tavares}}]{Agrawal:2020euj}%
  \BibitemOpen
  \bibfield  {author} {\bibinfo {author} {\bibfnamefont {P.}~\bibnamefont
  {Agrawal}}, \bibinfo {author} {\bibfnamefont {A.}~\bibnamefont {Hook}},
  \bibinfo {author} {\bibfnamefont {J.}~\bibnamefont {Huang}}, \ and\ \bibinfo
  {author} {\bibfnamefont {G.}~\bibnamefont {Marques-Tavares}},\ }\href@noop {}
  {\  (\bibinfo {year} {2020})},\ \Eprint {http://arxiv.org/abs/2010.15848}
  {arXiv:2010.15848 [hep-ph]} \BibitemShut {NoStop}%
\bibitem [{\citenamefont {Kim}(1979)}]{Kim:1979if}%
  \BibitemOpen
  \bibfield  {author} {\bibinfo {author} {\bibfnamefont {J.~E.}\ \bibnamefont
  {Kim}},\ }\href {\doibase 10.1103/PhysRevLett.43.103} {\bibfield  {journal}
  {\bibinfo  {journal} {Phys. Rev. Lett.}\ }\textbf {\bibinfo {volume} {43}},\
  \bibinfo {pages} {103} (\bibinfo {year} {1979})}\BibitemShut {NoStop}%
%%CITATION = PRLTA,43,103;%%
\bibitem [{\citenamefont {Shifman}\ \emph {et~al.}(1980)\citenamefont
  {Shifman}, \citenamefont {Vainshtein},\ and\ \citenamefont
  {Zakharov}}]{Shifman:1979if}%
  \BibitemOpen
  \bibfield  {author} {\bibinfo {author} {\bibfnamefont {M.~A.}\ \bibnamefont
  {Shifman}}, \bibinfo {author} {\bibfnamefont {A.~I.}\ \bibnamefont
  {Vainshtein}}, \ and\ \bibinfo {author} {\bibfnamefont {V.~I.}\ \bibnamefont
  {Zakharov}},\ }\href {\doibase 10.1016/0550-3213(80)90209-6} {\bibfield
  {journal} {\bibinfo  {journal} {Nucl. Phys.}\ }\textbf {\bibinfo {volume}
  {B166}},\ \bibinfo {pages} {493} (\bibinfo {year} {1980})}\BibitemShut
  {NoStop}%
%%CITATION = NUPHA,B166,493;%%
\bibitem [{\citenamefont {Carter}\ and\ \citenamefont
  {Martin}(1993)}]{Carter:1993wu}%
  \BibitemOpen
  \bibfield  {author} {\bibinfo {author} {\bibfnamefont {B.}~\bibnamefont
  {Carter}}\ and\ \bibinfo {author} {\bibfnamefont {X.}~\bibnamefont
  {Martin}},\ }\href {\doibase 10.1006/aphy.1993.1078} {\bibfield  {journal}
  {\bibinfo  {journal} {Annals Phys.}\ }\textbf {\bibinfo {volume} {227}},\
  \bibinfo {pages} {151} (\bibinfo {year} {1993})},\ \Eprint
  {http://arxiv.org/abs/hep-th/0306111} {arXiv:hep-th/0306111} \BibitemShut
  {NoStop}%
\bibitem [{\citenamefont {Brandenberger}\ \emph {et~al.}(1996)\citenamefont
  {Brandenberger}, \citenamefont {Carter}, \citenamefont {Davis},\ and\
  \citenamefont {Trodden}}]{Brandenberger:1996zp}%
  \BibitemOpen
  \bibfield  {author} {\bibinfo {author} {\bibfnamefont {R.~H.}\ \bibnamefont
  {Brandenberger}}, \bibinfo {author} {\bibfnamefont {B.}~\bibnamefont
  {Carter}}, \bibinfo {author} {\bibfnamefont {A.-C.}\ \bibnamefont {Davis}}, \
  and\ \bibinfo {author} {\bibfnamefont {M.}~\bibnamefont {Trodden}},\ }\href
  {\doibase 10.1103/PhysRevD.54.6059} {\bibfield  {journal} {\bibinfo
  {journal} {Phys. Rev. D}\ }\textbf {\bibinfo {volume} {54}},\ \bibinfo
  {pages} {6059} (\bibinfo {year} {1996})},\ \Eprint
  {http://arxiv.org/abs/hep-ph/9605382} {arXiv:hep-ph/9605382} \BibitemShut
  {NoStop}%
\bibitem [{\citenamefont {Martins}\ and\ \citenamefont
  {Shellard}(1998{\natexlab{a}})}]{Martins:1998gb}%
  \BibitemOpen
  \bibfield  {author} {\bibinfo {author} {\bibfnamefont {C.}~\bibnamefont
  {Martins}}\ and\ \bibinfo {author} {\bibfnamefont {E.}~\bibnamefont
  {Shellard}},\ }\href {\doibase 10.1103/PhysRevD.57.7155} {\bibfield
  {journal} {\bibinfo  {journal} {Phys. Rev. D}\ }\textbf {\bibinfo {volume}
  {57}},\ \bibinfo {pages} {7155} (\bibinfo {year} {1998}{\natexlab{a}})},\
  \Eprint {http://arxiv.org/abs/hep-ph/9804378} {arXiv:hep-ph/9804378}
  \BibitemShut {NoStop}%
\bibitem [{\citenamefont {Martins}\ and\ \citenamefont
  {Shellard}(1998{\natexlab{b}})}]{Martins:1998th}%
  \BibitemOpen
  \bibfield  {author} {\bibinfo {author} {\bibfnamefont {C.}~\bibnamefont
  {Martins}}\ and\ \bibinfo {author} {\bibfnamefont {E.}~\bibnamefont
  {Shellard}},\ }\href {\doibase 10.1016/S0370-2693(98)01466-X} {\bibfield
  {journal} {\bibinfo  {journal} {Phys. Lett. B}\ }\textbf {\bibinfo {volume}
  {445}},\ \bibinfo {pages} {43} (\bibinfo {year} {1998}{\natexlab{b}})},\
  \Eprint {http://arxiv.org/abs/hep-ph/9806480} {arXiv:hep-ph/9806480}
  \BibitemShut {NoStop}%
\bibitem [{\citenamefont {Carter}\ and\ \citenamefont
  {Davis}(2000)}]{Carter:1999an}%
  \BibitemOpen
  \bibfield  {author} {\bibinfo {author} {\bibfnamefont {B.}~\bibnamefont
  {Carter}}\ and\ \bibinfo {author} {\bibfnamefont {A.-C.}\ \bibnamefont
  {Davis}},\ }\href {\doibase 10.1103/PhysRevD.61.123501} {\bibfield  {journal}
  {\bibinfo  {journal} {Phys. Rev. D}\ }\textbf {\bibinfo {volume} {61}},\
  \bibinfo {pages} {123501} (\bibinfo {year} {2000})},\ \Eprint
  {http://arxiv.org/abs/hep-ph/9910560} {arXiv:hep-ph/9910560} \BibitemShut
  {NoStop}%
\bibitem [{\citenamefont {Barr}\ and\ \citenamefont
  {Matheson}(1987)}]{Barr:1987ij}%
  \BibitemOpen
  \bibfield  {author} {\bibinfo {author} {\bibfnamefont {S.~M.}\ \bibnamefont
  {Barr}}\ and\ \bibinfo {author} {\bibfnamefont {A.}~\bibnamefont
  {Matheson}},\ }\href {\doibase 10.1016/0370-2693(87)91486-9} {\bibfield
  {journal} {\bibinfo  {journal} {Phys. Lett. B}\ }\textbf {\bibinfo {volume}
  {198}},\ \bibinfo {pages} {146} (\bibinfo {year} {1987})}\BibitemShut
  {NoStop}%
\bibitem [{\citenamefont {Davis}\ \emph {et~al.}(2000)\citenamefont {Davis},
  \citenamefont {Perkins},\ and\ \citenamefont {Davis}}]{Davis:1999ec}%
  \BibitemOpen
  \bibfield  {author} {\bibinfo {author} {\bibfnamefont {S.~C.}\ \bibnamefont
  {Davis}}, \bibinfo {author} {\bibfnamefont {W.~B.}\ \bibnamefont {Perkins}},
  \ and\ \bibinfo {author} {\bibfnamefont {A.-C.}\ \bibnamefont {Davis}},\
  }\href {\doibase 10.1103/PhysRevD.62.043503} {\bibfield  {journal} {\bibinfo
  {journal} {Phys. Rev. D}\ }\textbf {\bibinfo {volume} {62}},\ \bibinfo
  {pages} {043503} (\bibinfo {year} {2000})},\ \Eprint
  {http://arxiv.org/abs/hep-ph/9912356} {arXiv:hep-ph/9912356} \BibitemShut
  {NoStop}%
\bibitem [{\citenamefont {Jeannerot}\ and\ \citenamefont
  {Postma}(2004)}]{Jeannerot:2004bs}%
  \BibitemOpen
  \bibfield  {author} {\bibinfo {author} {\bibfnamefont {R.}~\bibnamefont
  {Jeannerot}}\ and\ \bibinfo {author} {\bibfnamefont {M.}~\bibnamefont
  {Postma}},\ }\href {\doibase 10.1088/1126-6708/2004/12/032} {\bibfield
  {journal} {\bibinfo  {journal} {JHEP}\ }\textbf {\bibinfo {volume} {12}},\
  \bibinfo {pages} {032} (\bibinfo {year} {2004})},\ \Eprint
  {http://arxiv.org/abs/hep-ph/0411259} {arXiv:hep-ph/0411259} \BibitemShut
  {NoStop}%
\bibitem [{\citenamefont {Dreiner}\ \emph {et~al.}(2010)\citenamefont
  {Dreiner}, \citenamefont {Haber},\ and\ \citenamefont
  {Martin}}]{Dreiner:2008tw}%
  \BibitemOpen
  \bibfield  {author} {\bibinfo {author} {\bibfnamefont {H.~K.}\ \bibnamefont
  {Dreiner}}, \bibinfo {author} {\bibfnamefont {H.~E.}\ \bibnamefont {Haber}},
  \ and\ \bibinfo {author} {\bibfnamefont {S.~P.}\ \bibnamefont {Martin}},\
  }\href {\doibase 10.1016/j.physrep.2010.05.002} {\bibfield  {journal}
  {\bibinfo  {journal} {Phys. Rept.}\ }\textbf {\bibinfo {volume} {494}},\
  \bibinfo {pages} {1} (\bibinfo {year} {2010})},\ \Eprint
  {http://arxiv.org/abs/0812.1594} {arXiv:0812.1594 [hep-ph]} \BibitemShut
  {NoStop}%
\bibitem [{\citenamefont {Nielsen}\ and\ \citenamefont
  {Olesen}(1973)}]{Nielsen:1973cs}%
  \BibitemOpen
  \bibfield  {author} {\bibinfo {author} {\bibfnamefont {H.~B.}\ \bibnamefont
  {Nielsen}}\ and\ \bibinfo {author} {\bibfnamefont {P.}~\bibnamefont
  {Olesen}},\ }\href {\doibase 10.1016/0550-3213(73)90350-7} {\bibfield
  {journal} {\bibinfo  {journal} {Nucl. Phys.}\ }\textbf {\bibinfo {volume}
  {B61}},\ \bibinfo {pages} {45} (\bibinfo {year} {1973})}\BibitemShut
  {NoStop}%
%%CITATION = NUPHA,B61,45;%%
\bibitem [{\citenamefont {Hiramatsu}\ \emph {et~al.}(2011)\citenamefont
  {Hiramatsu}, \citenamefont {Kawasaki},\ and\ \citenamefont
  {Saikawa}}]{Hiramatsu:2010yn}%
  \BibitemOpen
  \bibfield  {author} {\bibinfo {author} {\bibfnamefont {T.}~\bibnamefont
  {Hiramatsu}}, \bibinfo {author} {\bibfnamefont {M.}~\bibnamefont {Kawasaki}},
  \ and\ \bibinfo {author} {\bibfnamefont {K.}~\bibnamefont {Saikawa}},\ }\href
  {\doibase 10.1088/1475-7516/2011/08/030} {\bibfield  {journal} {\bibinfo
  {journal} {JCAP}\ }\textbf {\bibinfo {volume} {08}},\ \bibinfo {pages} {030}
  (\bibinfo {year} {2011})},\ \Eprint {http://arxiv.org/abs/1012.4558}
  {arXiv:1012.4558 [astro-ph.CO]} \BibitemShut {NoStop}%
\bibitem [{\citenamefont {Hiramatsu}\ \emph {et~al.}(2012)\citenamefont
  {Hiramatsu}, \citenamefont {Kawasaki}, \citenamefont {Saikawa},\ and\
  \citenamefont {Sekiguchi}}]{Hiramatsu:2012gg}%
  \BibitemOpen
  \bibfield  {author} {\bibinfo {author} {\bibfnamefont {T.}~\bibnamefont
  {Hiramatsu}}, \bibinfo {author} {\bibfnamefont {M.}~\bibnamefont {Kawasaki}},
  \bibinfo {author} {\bibfnamefont {K.}~\bibnamefont {Saikawa}}, \ and\
  \bibinfo {author} {\bibfnamefont {T.}~\bibnamefont {Sekiguchi}},\ }\href
  {\doibase 10.1103/PhysRevD.85.105020} {\bibfield  {journal} {\bibinfo
  {journal} {Phys. Rev. D}\ }\textbf {\bibinfo {volume} {85}},\ \bibinfo
  {pages} {105020} (\bibinfo {year} {2012})},\ \bibinfo {note} {[Erratum:
  Phys.Rev.D 86, 089902 (2012)]},\ \Eprint {http://arxiv.org/abs/1202.5851}
  {arXiv:1202.5851 [hep-ph]} \BibitemShut {NoStop}%
\bibitem [{\citenamefont {Gorghetto}\ \emph {et~al.}(2018)\citenamefont
  {Gorghetto}, \citenamefont {Hardy},\ and\ \citenamefont
  {Villadoro}}]{Gorghetto:2018myk}%
  \BibitemOpen
  \bibfield  {author} {\bibinfo {author} {\bibfnamefont {M.}~\bibnamefont
  {Gorghetto}}, \bibinfo {author} {\bibfnamefont {E.}~\bibnamefont {Hardy}}, \
  and\ \bibinfo {author} {\bibfnamefont {G.}~\bibnamefont {Villadoro}},\ }\href
  {\doibase 10.1007/JHEP07(2018)151} {\bibfield  {journal} {\bibinfo  {journal}
  {JHEP}\ }\textbf {\bibinfo {volume} {07}},\ \bibinfo {pages} {151} (\bibinfo
  {year} {2018})},\ \Eprint {http://arxiv.org/abs/1806.04677} {arXiv:1806.04677
  [hep-ph]} \BibitemShut {NoStop}%
\bibitem [{\citenamefont {Gorghetto}\ \emph {et~al.}(2020)\citenamefont
  {Gorghetto}, \citenamefont {Hardy},\ and\ \citenamefont
  {Villadoro}}]{Gorghetto:2020qws}%
  \BibitemOpen
  \bibfield  {author} {\bibinfo {author} {\bibfnamefont {M.}~\bibnamefont
  {Gorghetto}}, \bibinfo {author} {\bibfnamefont {E.}~\bibnamefont {Hardy}}, \
  and\ \bibinfo {author} {\bibfnamefont {G.}~\bibnamefont {Villadoro}},\
  }\href@noop {} {\  (\bibinfo {year} {2020})},\ \Eprint
  {http://arxiv.org/abs/2007.04990} {arXiv:2007.04990 [hep-ph]} \BibitemShut
  {NoStop}%
\bibitem [{\citenamefont {Raffelt}(2008)}]{Raffelt:2006cw}%
  \BibitemOpen
  \bibfield  {author} {\bibinfo {author} {\bibfnamefont {G.~G.}\ \bibnamefont
  {Raffelt}},\ }\href {\doibase 10.1007/978-3-540-73518-2_3} {\bibfield
  {journal} {\bibinfo  {journal} {Lect. Notes Phys.}\ }\textbf {\bibinfo
  {volume} {741}},\ \bibinfo {pages} {51} (\bibinfo {year} {2008})},\ \Eprint
  {http://arxiv.org/abs/hep-ph/0611350} {arXiv:hep-ph/0611350} \BibitemShut
  {NoStop}%
\bibitem [{\citenamefont {Chang}\ \emph {et~al.}(2018)\citenamefont {Chang},
  \citenamefont {Essig},\ and\ \citenamefont {McDermott}}]{Chang:2018rso}%
  \BibitemOpen
  \bibfield  {author} {\bibinfo {author} {\bibfnamefont {J.~H.}\ \bibnamefont
  {Chang}}, \bibinfo {author} {\bibfnamefont {R.}~\bibnamefont {Essig}}, \ and\
  \bibinfo {author} {\bibfnamefont {S.~D.}\ \bibnamefont {McDermott}},\ }\href
  {\doibase 10.1007/JHEP09(2018)051} {\bibfield  {journal} {\bibinfo  {journal}
  {JHEP}\ }\textbf {\bibinfo {volume} {09}},\ \bibinfo {pages} {051} (\bibinfo
  {year} {2018})},\ \Eprint {http://arxiv.org/abs/1803.00993} {arXiv:1803.00993
  [hep-ph]} \BibitemShut {NoStop}%
\bibitem [{\citenamefont {Irastorza}\ and\ \citenamefont
  {Redondo}(2018)}]{Irastorza:2018dyq}%
  \BibitemOpen
  \bibfield  {author} {\bibinfo {author} {\bibfnamefont {I.~G.}\ \bibnamefont
  {Irastorza}}\ and\ \bibinfo {author} {\bibfnamefont {J.}~\bibnamefont
  {Redondo}},\ }\href {\doibase 10.1016/j.ppnp.2018.05.003} {\bibfield
  {journal} {\bibinfo  {journal} {Prog. Part. Nucl. Phys.}\ }\textbf {\bibinfo
  {volume} {102}},\ \bibinfo {pages} {89} (\bibinfo {year} {2018})},\ \Eprint
  {http://arxiv.org/abs/1801.08127} {arXiv:1801.08127 [hep-ph]} \BibitemShut
  {NoStop}%
\bibitem [{\citenamefont {Hamaguchi}\ \emph {et~al.}(2018)\citenamefont
  {Hamaguchi}, \citenamefont {Nagata}, \citenamefont {Yanagi},\ and\
  \citenamefont {Zheng}}]{Hamaguchi:2018oqw}%
  \BibitemOpen
  \bibfield  {author} {\bibinfo {author} {\bibfnamefont {K.}~\bibnamefont
  {Hamaguchi}}, \bibinfo {author} {\bibfnamefont {N.}~\bibnamefont {Nagata}},
  \bibinfo {author} {\bibfnamefont {K.}~\bibnamefont {Yanagi}}, \ and\ \bibinfo
  {author} {\bibfnamefont {J.}~\bibnamefont {Zheng}},\ }\href {\doibase
  10.1103/PhysRevD.98.103015} {\bibfield  {journal} {\bibinfo  {journal} {Phys.
  Rev. D}\ }\textbf {\bibinfo {volume} {98}},\ \bibinfo {pages} {103015}
  (\bibinfo {year} {2018})},\ \Eprint {http://arxiv.org/abs/1806.07151}
  {arXiv:1806.07151 [hep-ph]} \BibitemShut {NoStop}%
\bibitem [{\citenamefont {Weinberg}(1981)}]{Weinberg:1981eu}%
  \BibitemOpen
  \bibfield  {author} {\bibinfo {author} {\bibfnamefont {E.~J.}\ \bibnamefont
  {Weinberg}},\ }\href {\doibase 10.1103/PhysRevD.24.2669} {\bibfield
  {journal} {\bibinfo  {journal} {Phys. Rev. D}\ }\textbf {\bibinfo {volume}
  {24}},\ \bibinfo {pages} {2669} (\bibinfo {year} {1981})}\BibitemShut
  {NoStop}%
\bibitem [{\citenamefont {Jackiw}\ and\ \citenamefont
  {Rossi}(1981)}]{Jackiw:1981ee}%
  \BibitemOpen
  \bibfield  {author} {\bibinfo {author} {\bibfnamefont {R.}~\bibnamefont
  {Jackiw}}\ and\ \bibinfo {author} {\bibfnamefont {P.}~\bibnamefont {Rossi}},\
  }\href {\doibase 10.1016/0550-3213(81)90044-4} {\bibfield  {journal}
  {\bibinfo  {journal} {Nucl. Phys. B}\ }\textbf {\bibinfo {volume} {190}},\
  \bibinfo {pages} {681} (\bibinfo {year} {1981})}\BibitemShut {NoStop}%
\bibitem [{\citenamefont {Callan}\ and\ \citenamefont
  {Harvey}(1985)}]{Callan:1984sa}%
  \BibitemOpen
  \bibfield  {author} {\bibinfo {author} {\bibfnamefont {J.}~\bibnamefont
  {Callan}, \bibfnamefont {Curtis~G.}}\ and\ \bibinfo {author} {\bibfnamefont
  {J.~A.}\ \bibnamefont {Harvey}},\ }\href {\doibase
  10.1016/0550-3213(85)90489-4} {\bibfield  {journal} {\bibinfo  {journal}
  {Nucl. Phys. B}\ }\textbf {\bibinfo {volume} {250}},\ \bibinfo {pages} {427}
  (\bibinfo {year} {1985})}\BibitemShut {NoStop}%
\bibitem [{\citenamefont {Kaplan}\ and\ \citenamefont
  {Manohar}(1988)}]{Kaplan:1987kh}%
  \BibitemOpen
  \bibfield  {author} {\bibinfo {author} {\bibfnamefont {D.~B.}\ \bibnamefont
  {Kaplan}}\ and\ \bibinfo {author} {\bibfnamefont {A.}~\bibnamefont
  {Manohar}},\ }\href {\doibase 10.1016/0550-3213(88)90244-1} {\bibfield
  {journal} {\bibinfo  {journal} {Nucl. Phys. B}\ }\textbf {\bibinfo {volume}
  {302}},\ \bibinfo {pages} {280} (\bibinfo {year} {1988})}\BibitemShut
  {NoStop}%
\bibitem [{\citenamefont {Naculich}(1988)}]{Naculich:1987ci}%
  \BibitemOpen
  \bibfield  {author} {\bibinfo {author} {\bibfnamefont {S.~G.}\ \bibnamefont
  {Naculich}},\ }\href {\doibase 10.1016/0550-3213(88)90400-2} {\bibfield
  {journal} {\bibinfo  {journal} {Nucl. Phys. B}\ }\textbf {\bibinfo {volume}
  {296}},\ \bibinfo {pages} {837} (\bibinfo {year} {1988})}\BibitemShut
  {NoStop}%
\bibitem [{\citenamefont {Davis}(1988)}]{Davis:1988ip}%
  \BibitemOpen
  \bibfield  {author} {\bibinfo {author} {\bibfnamefont {R.~L.}\ \bibnamefont
  {Davis}},\ }\href {\doibase 10.1103/PhysRevD.38.3722} {\bibfield  {journal}
  {\bibinfo  {journal} {Phys. Rev. D}\ }\textbf {\bibinfo {volume} {38}},\
  \bibinfo {pages} {3722} (\bibinfo {year} {1988})}\BibitemShut {NoStop}%
\bibitem [{\citenamefont {Ringeval}(2001{\natexlab{a}})}]{Ringeval:2000kz}%
  \BibitemOpen
  \bibfield  {author} {\bibinfo {author} {\bibfnamefont {C.}~\bibnamefont
  {Ringeval}},\ }\href {\doibase 10.1103/PhysRevD.63.063508} {\bibfield
  {journal} {\bibinfo  {journal} {Phys. Rev. D}\ }\textbf {\bibinfo {volume}
  {63}},\ \bibinfo {pages} {063508} (\bibinfo {year} {2001}{\natexlab{a}})},\
  \Eprint {http://arxiv.org/abs/hep-ph/0007015} {arXiv:hep-ph/0007015}
  \BibitemShut {NoStop}%
\bibitem [{\citenamefont {Ringeval}(2001{\natexlab{b}})}]{Ringeval:2001xd}%
  \BibitemOpen
  \bibfield  {author} {\bibinfo {author} {\bibfnamefont {C.}~\bibnamefont
  {Ringeval}},\ }\href {\doibase 10.1103/PhysRevD.64.123505} {\bibfield
  {journal} {\bibinfo  {journal} {Phys. Rev. D}\ }\textbf {\bibinfo {volume}
  {64}},\ \bibinfo {pages} {123505} (\bibinfo {year} {2001}{\natexlab{b}})},\
  \Eprint {http://arxiv.org/abs/hep-ph/0106179} {arXiv:hep-ph/0106179}
  \BibitemShut {NoStop}%
\bibitem [{\citenamefont {Gouttenoire}\ \emph {et~al.}(2020)\citenamefont
  {Gouttenoire}, \citenamefont {Servant},\ and\ \citenamefont
  {Simakachorn}}]{Gouttenoire:2019kij}%
  \BibitemOpen
  \bibfield  {author} {\bibinfo {author} {\bibfnamefont {Y.}~\bibnamefont
  {Gouttenoire}}, \bibinfo {author} {\bibfnamefont {G.}~\bibnamefont
  {Servant}}, \ and\ \bibinfo {author} {\bibfnamefont {P.}~\bibnamefont
  {Simakachorn}},\ }\href {\doibase 10.1088/1475-7516/2020/07/032} {\bibfield
  {journal} {\bibinfo  {journal} {JCAP}\ }\textbf {\bibinfo {volume} {07}},\
  \bibinfo {pages} {032} (\bibinfo {year} {2020})},\ \Eprint
  {http://arxiv.org/abs/1912.02569} {arXiv:1912.02569 [hep-ph]} \BibitemShut
  {NoStop}%
\bibitem [{\citenamefont {Shifman}(2012)}]{Shifman:2012zz}%
  \BibitemOpen
  \bibfield  {author} {\bibinfo {author} {\bibfnamefont {M.}~\bibnamefont
  {Shifman}},\ }\href@noop {} {\emph {\bibinfo {title} {{Advanced topics in
  quantum field theory.}: {A lecture course}}}}\ (\bibinfo  {publisher}
  {Cambridge Univ. Press},\ \bibinfo {address} {Cambridge, UK},\ \bibinfo
  {year} {2012})\BibitemShut {NoStop}%
\bibitem [{\citenamefont {Zhitnitsky}(1980)}]{Zhitnitsky:1980tq}%
  \BibitemOpen
  \bibfield  {author} {\bibinfo {author} {\bibfnamefont {A.}~\bibnamefont
  {Zhitnitsky}},\ }\href@noop {} {\bibfield  {journal} {\bibinfo  {journal}
  {Sov. J. Nucl. Phys.}\ }\textbf {\bibinfo {volume} {31}},\ \bibinfo {pages}
  {260} (\bibinfo {year} {1980})}\BibitemShut {NoStop}%
\bibitem [{\citenamefont {Dine}\ \emph {et~al.}(1981)\citenamefont {Dine},
  \citenamefont {Fischler},\ and\ \citenamefont {Srednicki}}]{Dine:1981rt}%
  \BibitemOpen
  \bibfield  {author} {\bibinfo {author} {\bibfnamefont {M.}~\bibnamefont
  {Dine}}, \bibinfo {author} {\bibfnamefont {W.}~\bibnamefont {Fischler}}, \
  and\ \bibinfo {author} {\bibfnamefont {M.}~\bibnamefont {Srednicki}},\ }\href
  {\doibase 10.1016/0370-2693(81)90590-6} {\bibfield  {journal} {\bibinfo
  {journal} {Phys. Lett. B}\ }\textbf {\bibinfo {volume} {104}},\ \bibinfo
  {pages} {199} (\bibinfo {year} {1981})}\BibitemShut {NoStop}%
\end{thebibliography}%
%%%%%%%%%%%%%%%%%%%%%%%%%%%%%%%%%%%%%%%%%%%%

\end{document}